\documentclass[11pt,a4paper]{article}
\pdfoutput=1
\usepackage{jcappub}
\usepackage[T1]{fontenc}
\usepackage{lmodern}
\usepackage{makecell}
\usepackage{adjustbox}
\usepackage{placeins}
\hypersetup{hypertexnames=false,
  pdftitle={Spectral Constraints on Line-Seed Inverse-Compton Models for the MeV Feature in GRB 221009A},
  pdfauthor={Nissim Fraija}}
\apptocmd{\thebibliography}{\setlength{\itemsep}{3pt plus 0.2ex minus 0.05ex}}{}{}
\title{Spectral Constraints on Line-Seed Inverse-Compton Models for the MeV Feature in GRB 221009A}
\author{Nissim Fraija}
\affiliation{Instituto de Astronom\'ia, Universidad Nacional Aut\'onoma de M\'exico,\\
A. P. 70-264, Cd. Universitaria, Ciudad de M\'exico, M\'exico}
\emailAdd{nifraija@astro.unam.mx}

\abstract{We investigate inverse-Compton (IC) scattering of atomic and molecular line photons as an explanation of the evolving MeV spectral feature observed by Fermi-GBM in GRB~221009A. Detections of H$\alpha$ and Pa$\alpha$ emission from the host galaxy motivate hydrogen recombination seeds, with only indirect support for Ly$\alpha$. A marginal Mg~XII feature in the late-time X-ray afterglow provides an event-specific motivation for metal-line seeds, whereas Si~XIV, S~XVI and iron K-shell transitions are motivated by historical reports of X-ray line features in other gamma-ray bursts. Strong late-time H$_2$ emission detected by the James Webb Space Telescope at the projected burst position motivates our most detailed spectral test. Assuming lateral illumination by the H$_2$ $1-0\,S(1)$ transition at $2.122\,\mu{\rm m}$, we calibrate lognormal electron distributions to reproduce the observed MeV photon centroids and widths. We also require IC emission from the ultraviolet (UV) photons driving molecular fluorescence to remain compatible with contemporaneous LAT observations. In a separate test including accompanying molecular transitions, the most favourable configuration, based on the Rh3o photodissociation-region model, approximately describes the MeV feature, although all tested configurations exceed the adopted central spectral references at some energies. For an assumed precursor UV budget, favourable interception substantially reduces the seed-photon energy deficit, but even under ideal excitation and internal escape assumptions, the reference-line photon density in the compact examples falls short of the required value by factors of approximately 4--9. Collisional de-excitation, infrared photon escape and electron cooling impose additional constraints on these configurations. Atomic candidates remain kinematically possible, but their local seed-photon supply is also unestablished.}

\keywords{gamma ray bursts theory, absorption and radiation processes, gamma ray theory}

\begin{document}
\maketitle
\flushbottom

\section{Introduction}

Spectral features in gamma-ray bursts (GRBs) can constrain the radiating particles and their environment beyond what is possible from a smooth continuum alone. Line production in GRB environments has observational and theoretical precedents. Possible iron K-shell features were reported in the X-ray afterglows of GRB~970508 and GRB~991216 \cite{Piro1999,Piro2000}, while features attributed to Si~XIV and S~XVI were claimed in GRB~011211 \cite{Reeves2002}. These reports motivated models of irradiated circumburst material \cite{2002ApJ...572L..57L,KumarNarayan2003} and provide historical motivation for considering these transitions as candidate atomic seeds here. The statistical significance and physical interpretation of these features have remained debated \cite{Bottcher2004,Sako2005}.

GRB~221009A, at a redshift of $z=0.151$ \cite{2025A&A...701A.134M}, provides a particularly useful case for investigating spectral features within an exceptionally bright, evolving gamma-ray spectrum \cite{2025ApJS..277...24A}. Using data from the Fermi Gamma-ray Burst Monitor (GBM), Ravasio et al.~\cite{2024Sci...385..452R} reported a significant MeV emission feature whose characteristic energy and luminosity decreased with time, interpreting it as Doppler-boosted electron--positron annihilation radiation. Subsequent analyses examined its temporal evolution \cite{2024SCPMA..6789511Z}, while an independent analysis by the Fermi-GBM team supported its statistical significance and found no instrumental explanation for the signal \cite{2024arXiv241000286B}. These results have renewed interest in line formation during the prompt emission of GRBs.

Complementing these studies, Liao et al.~\cite{2025ApJ...983..130L} searched for spectral lines in the 30 highest-fluence GRBs in the Fermi/GBM catalogue. They confirmed no additional astrophysical lines and identified recurring features consistent with instrumental effects. Their findings emphasize the importance of testing instrumental effects and the adopted continuum description when assessing the physical origin of a spectral feature.

Several studies have examined the physical requirements of the pair-annihilation interpretation. Yi et al.~\cite{2024MNRAS.535..982Y} used optical-depth arguments to constrain the regions in which pairs could be produced, cool and annihilate while allowing the resulting radiation to escape. Salafia et al.~\cite{2026A&A...712A..97S} developed a scenario in which high-latitude emission from a thin shell of annihilating pairs accounts for the observed line properties under specific constraints on the shell radius, Lorentz factor and pair content. A different interpretation was proposed by Moradi et al.~\cite{2026CmPhy...9..172M}, who associated the evolving MeV feature with Doppler-boosted radioactive decay of freshly synthesized $^{56}$Ni entrained in the relativistic outflow. These proposals connect the feature to different aspects of the emitting environment and motivate comparisons based on both spectral properties and physical consistency.

Inverse-Compton (IC) scattering offers a possible connection between lower-energy seed photons and a spectral feature at MeV energies. In the Thomson regime, the scattered-photon energy scales approximately as the seed energy multiplied by the square of the electron Lorentz factor, with additional dependence on the incident and outgoing directions and cosmological redshift \cite{Blumenthal1970,2010ApJS..189..286P}. This scaling alone does not determine the seed species: different seed energies can be compensated by different electron energies and geometries. A physically meaningful comparison must also examine the spectral width generated by the scattering kernel, the photon density encountered by the electrons, and the line and continuum emission associated with the seed source. The full scattering kinematics must be retained when recoil becomes important.

Observations of GRB~221009A and its host galaxy provide additional motivation for investigating atomic and molecular seed photons. H$\alpha$ and Pa$\alpha$ emission has been detected in the host galaxy \cite{2025A&A...701A.134M}, while a marginal feature consistent with Mg~XII emission was reported in the X-ray afterglow approximately 2.3 days after the burst \cite{2024ApJ...972...75C}. Late-time observations with the James Webb Space Telescope (JWST) also revealed strong H$_2$ emission at the projected GRB position, with line ratios favouring ultraviolet (UV) fluorescence \cite{2024NatAs...8..774B}. These findings motivate examining hydrogen recombination lines, metal transitions and molecular emission as possible IC seeds, including the UV radiation responsible for molecular excitation. However, the observed host and afterglow emission does not establish the intensity, geometry or temporal availability of the local photon fields required during the MeV episode.

Against this background, we investigate atomic and molecular line photons as seeds for an IC interpretation of the MeV feature in GRB~221009A.   For the H$_2$--UV scenario, we calibrate electron distributions
to reproduce the observed MeV photon centroids and widths
and require the accompanying UV-scattered emission to remain
compatible with the published spectral models of contemporaneous
Fermi Large Area Telescope (LAT) emission
\cite{2025ApJS..277...24A}.
Our principal quantitative test separately examines whether
the accompanying molecular transitions produce excess emission
relative to the adopted MeV spectral reference after normalizing
the molecular IC spectrum to the observed feature.
This test covers a prescribed set of molecular spectra,
attenuations and widths.  We also explore the physical requirements for supplying the seed radiation, including its production, escape and geometrical distribution, and consider precursor-assisted molecular excitation as a possible preparation mechanism. The resulting constraints apply to the configurations examined here and do not constitute a statistical exclusion of line-seed IC scattering in general. Section~\ref{sec:observations} describes the observational constraints; Section~\ref{ref:scattering} defines the scattering model; Section~\ref{sec:seeds} introduces the seed candidates; Section~\ref{sec:discussion} presents their requirements and the molecular consistency test; and Section~\ref{sec:summary} summarises the conclusions.

\section{Observational constraints on GRB 221009A}
\label{sec:observations}

GRB~221009A triggered Fermi-GBM on 2022 October 9 at
13:16:59.99~UTC. We adopt this epoch as $T_0$, following
Axelsson et al.~\cite{2025ApJS..277...24A}, and express observing intervals
relative to it. We adopt a redshift of $z=0.151$
\cite{2025A&A...701A.134M}. The burst had a total isotropic-equivalent radiated energy
of $E_{\gamma,\mathrm{iso}}\simeq10^{55}\,\mathrm{erg}$,
derived by Lesage et al.~\cite{2023ApJ...952L..42L} from their $k$-corrected
1--10\,000~keV analysis over
$T_0+[-2.7,1449.5]$~s. They also reported a one-second
peak isotropic-equivalent luminosity of
$L_{\gamma,\mathrm{iso,pk}}\simeq
9.9\times10^{53}\,\mathrm{erg\,s^{-1}}$.
These quantities provide the overall energetic context
for assessing the MeV spectral feature. Photon energies and time intervals
are expressed in the observer frame unless otherwise stated;
quantities referring to the host-galaxy or plasma comoving
frame are identified explicitly.

Our observational inputs are published spectral parameters
and model spectra. For the quantitative comparison, we
consider $T_0+[290.6,300.6]$~s and
$T_0+[300.6,325.6]$~s, adopting the Gaussian MeV-component
parameters and Fermi-LAT power-law reference spectra reported
by Axelsson et al.~\cite{2025ApJS..277...24A}. Their normalization and temporal
matching are described in Section~\ref{H2-UV_scenario}. These inputs constrain
the characteristic energy, width and flux of the MeV feature
and provide a reference for assessing accompanying
high-energy emission. Late-time JWST spectroscopy supplies
the observational motivation for molecular seeds, as
discussed below.

Ravasio et al.~\cite{2024Sci...385..452R} reported an emission feature whose centroid decreased from approximately 12 to 6 MeV over an interval of about 80 s. A Gaussian description yielded a characteristic width of order
1~MeV. The line luminosity was approximately
$1.1\times10^{50}\,{\rm erg\,s^{-1}}$ in
$T_0+[280,300]$~s, whereas a $2\sigma$ upper limit of
$4.3\times10^{49}\,{\rm erg\,s^{-1}}$ was obtained for
$T_0+[340,360]$~s \cite{2024Sci...385..452R}. These values provide representative scales for our initial estimates. The Gaussian width refers to the standard deviation, $\sigma_E$, rather than the full width at half maximum, which is $2\sqrt{2\ln 2}\,\sigma_E$ \cite{2024Sci...385..452R}.

A viable scattering scenario must therefore produce a localized spectral enhancement that shifts towards lower energies while fading. Matching a representative centroid is insufficient: the model must also account for the finite width and its evolution. We characterize the relative width through $w=\sigma_E/E_{\rm c}$, where $E_{\rm c}$ denotes the feature centroid. Our illustrative calibration uses the published central values
of the centroid, width and normalization for each observing
interval. Their uncertainties are not propagated through this
calculation, and the comparison remains conditional on the
adopted continuum description.

The LAT observations provide a reference for assessing whether
the accompanying emission from IC scattering of ultraviolet
seed photons (IC--UV) in the molecular scenario overproduces
the contemporaneous high-energy emission. Axelsson et al.~\cite{2025ApJS..277...24A} characterized the temporal and spectral properties of GRB 221009A at LAT energies, accounting for the instrumental effects associated with its exceptional brightness. We use the published spectral models as references for the accompanying emission predicted by a common electron population. This comparison does not identify a distinct LAT component associated with the MeV feature. In particular, it is not an independent event-level LAT likelihood analysis and yields neither a detection significance nor a statistical upper limit on the normalization of this component.  The LAT comparison therefore tests the spectral compatibility
of the predicted contribution, without requiring it to account
for all of the observed high-energy emission.

Late-time JWST spectroscopy provides observational motivation for considering molecular seed photons. Blanchard et al.~\cite{2024NatAs...8..774B} identified narrow $H_2$ emission lines that are particularly strong at the projected explosion site of GRB 221009A. Their comparison of the measured line ratios with excitation models favoured fluorescence over shock excitation, motivating an investigation of molecular emission together with the UV radiation responsible for its excitation. These observations establish the presence of line-emitting molecular gas in the projected environment of the burst. However, they do not directly determine the photon density, angular distribution or temporal availability at the scattering location during the prompt MeV episode.

This distinction applies to all atomic and molecular seed
candidates considered here. For each candidate, the emitting
region must supply the required photon density and illumination
geometry during the MeV episode. Observational evidence for
a transition therefore motivates its consideration but does
not, by itself, establish the local radiation field required
by the IC model.

\section{Inverse-Compton model}
\label{ref:scattering}

We formulate the IC framework used to assess candidate seed-photon fields. We specify the reference frames, scattering geometry, and particle distributions, including recoil and cosmological redshift. This framework relates the seed radiation to the energy, width and flux of the scattered emission, providing the basis for comparison with the MeV feature and contemporaneous LAT emission.

\subsection{Reference frames and scattering kinematics}

We consider IC scattering of an external radiation field by relativistic electrons. The seed field may contain both discrete transitions and accompanying continuum emission.  We first describe the kinematics of a single scattering event; the observable spectrum subsequently requires integration over the electron and photon distributions and the emitting geometry \cite{Blumenthal1970, 2010ApJS..189..286P}.

We use three reference frames. Unstarred quantities, including the seed photon energy $\epsilon$, electron Lorentz factor $\gamma_e$, positions and directions, are measured in the host-galaxy frame, in which the candidate seed sources are taken to be stationary. Starred quantities are measured in the rest frame of an individual scattering electron.
The subscript ``obs'' denotes quantities measured by the distant observer. For photon energies, the corresponding host-frame value is larger by a factor $(1+z)$.  In the host-galaxy frame, an electron has Lorentz factor $\gamma_e$ and velocity $\beta_e c\,\hat{\boldsymbol e}$, where
$\hat{\boldsymbol e}$ is the unit vector along the electron velocity and $\beta_e=(1-\gamma_e^{-2})^{1/2}$.

An incident photon has energy $\epsilon$ and direction $\hat{\boldsymbol n}_i$, while the scattered photon propagates towards the observer along $\hat{\boldsymbol n}_s$. We define
$\mu_i=\hat{\boldsymbol e}\cdot\hat{\boldsymbol n}_i$,
$\mu_s=\hat{\boldsymbol e}\cdot\hat{\boldsymbol n}_s$, and
$\cos\psi=\hat{\boldsymbol n}_i\cdot\hat{\boldsymbol n}_s$. 
An incident seed photon has host-frame energy $\epsilon_i\equiv\epsilon$, whereas the energy of the IC-scattered photon in the host and observer
frames is denoted by $\epsilon^{\rm IC}$ and
$\epsilon_{\rm obs}^{\rm IC}$, respectively. Energy–momentum conservation gives the observed scattered-photon energy as \citep[e.g.][]{Blumenthal1970, 2010ApJS..189..286P}

\begin{eqnarray}
\epsilon^{\rm IC}_{\rm obs}=
\frac{\epsilon}{1+z}
\frac{1-\beta_e\mu_{\rm i}}
{1-\beta_e\mu_{\rm s}
+\dfrac{\epsilon}{\gamma_e m_ec^2}(1-\cos\psi)}.
\label{eq:E_obs}
\end{eqnarray}

Here, $\gamma_e$ is measured in the host-galaxy
frame; it is therefore not, in general, the random Lorentz factor in the plasma comoving frame. The incident photon energy in the electron rest frame is

\begin{eqnarray}
\epsilon_\star=\gamma_e\epsilon(1-\beta_e\mu_{\rm i}).
\label{eq:ph_i}
\end{eqnarray}

In the Thomson limit, \(\epsilon_\star\ll m_ec^2\), recoil can be neglected. For an ultrarelativistic electron directed towards the observer, \(\mu_{\rm s}=1\), equation (\ref{eq:E_obs}) then reduces to

\begin{eqnarray}
\epsilon^{\rm IC}_{\rm obs}\simeq
\frac{2\gamma_e^2\epsilon}{1+z}
(1-\beta_e\mu_{\rm i}).
\label{eq:E_obs_s}
\end{eqnarray}

Consequently, the energy amplification depends on both the electron energy and the incidence angle. Head-on photons yield $\epsilon^{\rm IC}_{\rm obs}\simeq4\gamma_e^2\epsilon/(1+z)$,
whereas exactly co-propagating photons observed along the electron velocity
retain their host-frame energy, giving
$\epsilon^{\rm IC}_{\rm obs}=\epsilon/(1+z)$.

In particular, for two seed ($\epsilon_{\rm 1}$, $\epsilon_{\rm 2}$) components   scattered by the same electrons in the Thomson regime, their characteristic IC-scattered energies approximately satisfy

\begin{eqnarray}
\frac{\epsilon^{\rm IC}_{\rm obs, 2}}{\epsilon^{\rm IC}_{\rm obs, 1}}\simeq
\frac{\epsilon_{\rm 2}}{\epsilon_{\rm 1}}
\frac{1-\beta_e\mu_{{\rm i},2}}
     {1-\beta_e\mu_{{\rm i},1}},
\label{eq:component-energy-ratio}
\end{eqnarray}

provided that the components sample comparable electron energies and outgoing directions. This relation provides an initial kinematic test of whether a seed line
and accompanying radiation can produce components in different observed energy bands. It does not determine their relative luminosities, which depend on the photon densities and angular distributions, the electron population and the scattering cross-section. 
%For identical incidence angles, equation~(\ref{eq:component-energy-ratio}) reduces to the ratio of seed-photon energies. This limiting case will be used only as an energy diagnostic; the directional calculation below retains the distinct angular distributions and recoil corrections.

For head-on incidence and scattering along the velocity of an ultrarelativistic electron, retaining recoil in equation~(\ref{eq:E_obs}) gives $\epsilon_{\rm obs}^{\rm IC}\simeq4\gamma_e^2\epsilon/[(1+z)(1+b)]$, where $b\equiv4\gamma_e\epsilon/(m_ec^2)$ is the dimensionless recoil parameter. In the strong-recoil limit, $b\gg1$, the scattered-photon energy approaches $\gamma_e m_ec^2/(1+z)$. At fixed electron energy and scattering geometry, the response to a change in seed-photon energy is

\begin{eqnarray}
\frac{d\ln \epsilon^{\rm IC}_{\rm obs}}{d\ln\epsilon}
=\frac{1}{1+b}.
\label{eq:seed_energy_response}
\end{eqnarray}

This relation applies to both atomic and molecular seeds under the stated assumptions. For $b\ll1$, a small fractional seed-energy spread is transmitted almost unchanged, whereas recoil reduces its contribution to the scattered-energy spread. This is a local response at fixed electron energy and angles, not a prediction of the full spectral width.

These relations describe single-scattering kinematics rather than an emergent line profile. Even monochromatic seeds can produce a broad scattered component when the incident angles or electron energies span a substantial range. The spectral width and temporal evolution must therefore be calculated from the full distributions.

\subsection{Electron and seed-photon distributions}

We describe the scattering particles and seed radiation locally in the
host-galaxy frame. Let $f_e(\gamma_e,\Omega_e,\boldsymbol r,t)$ denote the electron number density per unit Lorentz factor and solid angle, where $\Omega_e$ is the solid angle specifying the direction of the
electron velocity $\hat{\boldsymbol e}$.  We write

\begin{eqnarray}
f_e(\gamma_e,\Omega_e,\boldsymbol r,t)
=
n_e(\boldsymbol r,t)\,
g_e(\gamma_e,\Omega_e;\boldsymbol r,t),
\qquad
\int_1^\infty d\gamma_e\int d\Omega_e\,g_e=1\,.
\end{eqnarray}

Here, $n_e(\boldsymbol r,t)$ is the local electron number density and $g_e(\gamma_e,\Omega_e;\boldsymbol r,t)$ is the normalized distribution of electron Lorentz factors and velocity directions.  For the reference calculations developed below, we adopt a monoenergetic and unidirectional electron distribution. Thus, $g_e=\delta_{\rm D}(\gamma_e-\gamma_0)
\delta^{(2)}(\Omega_e-\Omega_0)$, where $\delta_{\rm D}$ is the one-dimensional Dirac distribution and
$\delta^{(2)}$ is the Dirac distribution on the two-dimensional angular
space. Writing $\Omega_e=(\theta_e,\phi_e)$, with
$d\Omega_e=d(\cos\theta_e)\,d\phi_e$, it is defined by $\delta^{(2)}(\Omega_e-\Omega_0)=\delta(\cos\theta_e-\cos\theta_0)\, \delta(\phi_e-\phi_0)$, so that $\int d\Omega_e\,\delta^{(2)}(\Omega_e-\Omega_0)=1$.
This is a reference limiting case that isolates the broadening caused by
the seed radiation and scattering geometry. The time-resolved calculation in Section~5.4 relaxes this assumption and adopts a finite lognormal electron distribution. We define the seed-photon distribution as

\begin{eqnarray}
n_{\rm ph}(\epsilon,\Omega_{\rm i},\boldsymbol r,t)
\equiv
\frac{dN_{\rm ph}}{dV\,d\epsilon\,d\Omega_{\rm i}}
=
\sum_j n_j(\boldsymbol r,t)\,
\Phi_j(\epsilon,\Omega_{\rm i};\boldsymbol r,t)
+
n_{\rm cont}(\epsilon,\Omega_{\rm i},\boldsymbol r,t),
\end{eqnarray}

where $dV$ is an infinitesimal volume element centred at
$\boldsymbol r$, $d\epsilon$ is a seed-photon energy interval, and
$d\Omega_i$ is an element of incident-photon solid angle. Thus,
$n_{\rm ph}$ is the differential photon number density per unit volume,
photon energy and solid angle. The quantity $n_j(\boldsymbol r,t)$ is the
local photon number density integrated over the seed line $j$, whereas
$\Phi_j(\epsilon,\Omega_i;\boldsymbol r,t)$ is its normalized joint
energy--angular distribution, with units of energy$^{-1}$\,sr$^{-1}$.
Each line profile is normalized such that
$\int_0^\infty d\epsilon\int d\Omega_i\,\Phi_j=1$.  The term $n_{\rm cont}(\epsilon,\Omega_i,\boldsymbol r,t)$ is the
differential photon number density of the accompanying continuum per unit photon energy and solid angle, and therefore has the same units as $n_{\rm ph}$. It includes radiation associated with the line-production mechanism and any additional seed component explicitly included in the
model.

For each line, the local radiation energy density is
\begin{eqnarray}
u_j(\boldsymbol r,t)=n_j(\boldsymbol r,t)\,\bar{\epsilon}_j,
\qquad
\bar{\epsilon}_j=
\int_0^\infty\int
\epsilon\,\Phi_j(\epsilon,\Omega_i;\boldsymbol r,t)\,
d\epsilon\,d\Omega_i .
\end{eqnarray}

Here, $\bar{\epsilon}_j$ is the mean host-frame photon energy of seed line $j$. Thus, the energy density is the local line-photon number density multiplied by its mean photon energy. These local densities determine the scattering rate and must be related to a physical source geometry. An observed line luminosity alone does not specify the radiation density or incident directions at the electron location.

The LAT consistency test applies specifically to the H$_2$--UV scenario, because the electrons scattering the molecular line also upscatter the associated UV radiation. The other seed candidates are assessed as possible sources of the MeV feature, without requiring an accompanying LAT component.

The monoenergetic assumption does not by itself guarantee a narrow scattered feature. The emergent profile also depends on the seed-line width, the distribution of incidence angles, the angular scattering probability and integration over the observing interval. These contributions must be retained when testing each candidate.

\subsection{Anisotropic Inverse-Compton Emissivity}
\label{sec:ic_emissivity}
We calculate the local emission in the single-scattering approximation, assuming unpolarized seed photons and negligible magnetic modifications to the scattering cross-section at the interaction site. The latter assumption concerns the scattering region and does not preclude a magnetized source of seed photons. Electron recoil is retained through the Klein–Nishina cross-section.

Let $\Omega_s$ specify the outgoing photon direction. We define the differential photon
production rate as

\begin{eqnarray}
q_{\rm IC}(\epsilon^{\rm IC},\Omega_s,\boldsymbol r,t)
&\equiv&
\frac{dN_{\rm sc}}
{dt\,dV\,d\epsilon^{\rm IC}\,d\Omega_s} \nonumber\\
&=&
c\int d\gamma_e\,d\Omega_e\,
f_e(\gamma_e,\Omega_e,\boldsymbol r,t)
\int d\epsilon\,d\Omega_i\,
n_{\rm ph}(\epsilon,\Omega_i,\boldsymbol r,t)
(1-\beta_e\mu_i)\,\mathcal K .
\end{eqnarray}

Here, $N_{\rm sc}$ is the number of scattered photons and $t$ is the
host-frame time. All distributions are evaluated at the same position and
time. The factor $c(1-\beta_e\mu_i)$ accounts for the photon--electron
encounter rate, while $\mathcal K$ specifies the angular probability and
scattered energy for each encounter.

For a specified electron velocity and incident and outgoing photon directions, the redistribution kernel can be written as
\begin{eqnarray}
\mathcal K=
\frac{d\sigma_\star}{d\Omega_{{\rm s},\star}}\,
\frac{1}{\gamma_e^2(1-\beta_e\mu_{\rm s})^2}\,
\delta_{\rm D}\left[\epsilon^{\rm IC}-
\frac{\epsilon(1-\beta_e\mu_{\rm i})}
{1-\beta_e\mu_{\rm s}
+\dfrac{\epsilon}{\gamma_e m_ec^2}(1-\cos\psi)}\right].
\end{eqnarray}

 The factor preceding the Dirac distribution includes the solid-angle Jacobian, $\frac{d\Omega_{{\rm s},\star}}{d\Omega_{\rm s}}=\frac{1}{\gamma_e^2(1-\beta_e\mu_{\rm s})^2}$.  The Dirac distribution enforces the energy of a single scattering event; integration over the incident radiation and electron distributions generates the emergent spectral profile.

For unpolarized radiation, the electron-rest-frame differential
Klein--Nishina cross-section is \cite{KleinNishina1929}

\begin{eqnarray}
\frac{d\sigma_\star}{d\Omega_{{\rm s},\star}}
=
\frac{3\sigma_{\rm T}}{16\pi}\,
\eta^2
\left(
\eta+\eta^{-1}-\sin^2\Theta_\star
\right),
\qquad
\eta=
\left[
1+\frac{\epsilon_\star}{m_ec^2}
(1-\cos\Theta_\star)
\right]^{-1}\,.
\end{eqnarray}

Here, \(\sigma_{\rm T}\) is the Thomson cross-section, \(\eta\equiv\epsilon^{\rm IC}_\star/\epsilon_\star\) is the ratio of the scattered to incident photon energies in the electron rest frame, and \(\Theta_\star\) is the angle between the incident and scattered photons in that frame.  The required angle follows from

\begin{eqnarray}
1-\cos\Theta_\star
=
\frac{1-\cos\psi}
{\gamma_e^2(1-\beta_e\mu_{\rm i})(1-\beta_e\mu_{\rm s})}.
\end{eqnarray}

 These expressions retain the angular dependence that is essential for directional seed radiation; anisotropic Compton redistribution is discussed by \cite{2010ApJS..189..286P}.

For prescribed electron and photon distributions, the emissivity is linear in the seed components, $q_{\rm IC}=\sum_j q_{{\rm IC},j} +q_{{\rm IC},{\rm cont}}$. Thus, the candidate line and accompanying radiation are scattered by the same electron population using a common kernel. Their relative contributions to GBM and LAT depend on the seed densities, incidence angles and recoil corrections. This linear decomposition applies to the instantaneous emissivity; strong radiative cooling or Compton drag can change the electron distribution and must be included when calculating its evolution.

The corresponding energy emissivity is \(j_{\rm IC}=\epsilon^{\rm IC}q_{\rm IC}\). Obtaining the observed spectrum requires integration over the emitting region at equal photon arrival time, followed by the cosmological transformation and any relevant attenuation. The single-scattering approximation must also be checked against the optical depth of each proposed configuration.

\subsection{Observed flux and temporal evolution}
\label{sec:Obs_flux_evol}

The observed signal combines radiation emitted at different positions and host-frame times. For a distant observer in direction \(\hat{\boldsymbol n}_{\rm s}\), the photon arrival time satisfies \cite[e.g.][]{1999astro.ph..5116H, 1999ApJ...513..679G}

\begin{eqnarray}
t_{\rm obs}=(1+z)
\left(t-\frac{\hat{\boldsymbol n}_{\rm s}\cdot\boldsymbol r}{c}\right)
+t_{\rm off},
\end{eqnarray}

where \(t_{\rm off}\) fixes the observational time origin. Thus, the emissivity contributing at a given arrival time must be evaluated at

\begin{eqnarray}
t_{\rm ret}(\boldsymbol r,t_{\rm obs})
=
\frac{t_{\rm obs}-t_{\rm off}}{1+z}
+\frac{\hat{\boldsymbol n}_{\rm s}\cdot\boldsymbol r}{c}.
\end{eqnarray}

Here, \(t_{\rm ret}(\boldsymbol r,t_{\rm obs})\) is the retarded host-frame time at which radiation emitted from position \(\boldsymbol r\) arrives at the observer at \(t_{\rm obs}\). Let \(F_{\epsilon_{\rm obs}}\) denote the observed energy flux per unit photon energy.
Using the host-frame emissivity per unit solid angle defined
in Section~\ref{sec:ic_emissivity} and the cosmological spectral flux--luminosity relation \cite{1999astro.ph..5116H}, we obtain

\begin{eqnarray}
F_{\epsilon_{\rm obs}}(\epsilon_{\rm obs},t_{\rm obs})
=
\frac{1+z}{d_L^2}
\int d^3\boldsymbol r\,
j_{\rm IC}\!\left[
(1+z)\epsilon_{\rm obs},
\Omega_{\rm s},
\boldsymbol r,t_{\rm ret}
\right]
e^{-\tau_{\rm esc}},
\end{eqnarray}

where \(d_L\) is the luminosity distance and \(\tau_{\rm esc}\) is the optical depth for removal of photons from the direct signal along the outgoing ray. The integration uses host-frame spatial coordinates and includes the full evolving source through \(j_{\rm IC}\). No additional Doppler factor is applied, since the scattering kinematics and angular transformation are already included in that emissivity. Likewise, no factor \(1/(4\pi)\) is required because \(j_{\rm IC}\) is defined per unit solid angle.

The attenuation factor describes the directly escaping component. If absorption produces a significant secondary cascade, its emission must be calculated separately. The optically thin limit corresponds to \(\tau_{\rm esc}\ll1\).

For an observing interval \([t_a,t_b]\), the time-averaged spectrum is

\begin{eqnarray}
\overline{F}_{\epsilon_{\rm obs}}(\epsilon_{\rm obs})
=
\frac{1}{t_b-t_a}
\int_{t_a}^{t_b}
F_{\epsilon_{\rm obs}}(\epsilon_{\rm obs},t_{\rm obs})\,dt_{\rm obs}.
\end{eqnarray}

Both the MeV feature and the predicted LAT component are evaluated over matched intervals. Integration over time is essential because an evolving centroid can broaden the measured feature even when the instantaneous spectrum is narrow. When detector data are fitted directly, the predicted photon flux, \(F_{\epsilon_{\rm obs}}/\epsilon_{\rm obs}\), must be folded through the relevant instrumental response.

The arrival-time relation also connects the observed duration to the distance travelled by the scattering electrons. Along a trajectory \(d\boldsymbol r/dt=\beta_e c\,\hat{\boldsymbol e}\),

\begin{eqnarray}
dt_{\rm obs}
=
(1+z)(1-\beta_e\mu_{\rm s})\,dt.
\end{eqnarray}

For approximately constant speed and direction, a travelled distance \(\Delta\ell\) therefore gives $\Delta t_{\rm obs} = (1+z) \frac{1-\beta_e\mu_{\rm s}}{\beta_e c}\,\Delta\ell$.

For ultrarelativistic motion towards the observer, this reduces to

\begin{eqnarray}
\Delta\ell\simeq
\frac{2\gamma_e^2c\,\Delta t_{\rm obs}}{1+z}.
\end{eqnarray}

This is a trajectory length and need not equal the distance from the central engine or the size of the seed source.

Finally, the temporal evolution of the scattered feature can arise from changes in electron energy, seed-photon incidence angles or the seed spectrum encountered along the trajectory. In the aligned Thomson limit,

\begin{eqnarray}
\epsilon^{\rm IC}_{\rm obs}(t_{\rm obs})
\simeq
\frac{2\gamma_e^2(t)\epsilon(t)}{1+z}
\left[1-\beta_e(t)\mu_{\rm i}(t)\right],
\end{eqnarray}

with host-frame quantities evaluated at the corresponding scattering event. Consequently, a decreasing centroid does not uniquely imply electron deceleration. The accompanying luminosity and LAT evolution provide additional constraints on these possibilities.

\subsection{Seed-photon requirements and energy budget}

The seed-photon density required to reproduce the MeV feature must be consistent with the luminosity and geometry of its proposed source. For a compact, stationary source emitting line photons isotropically, free propagation gives

\begin{eqnarray}
u_j(\boldsymbol r,t)=
\frac{L_j(t-d/c)}{4\pi c\,d^2},
\qquad
n_j(\boldsymbol r,t)=
\frac{L_j(t-d/c)}
{4\pi c\,d^2\bar{\epsilon}_j},
\end{eqnarray}

where \(d\) is the distance between the seed source and the scattering location, and \(L_j\) is the line luminosity. These expressions assume negligible attenuation between the two locations. Extended sources require integration over their spatial emissivity, while absorption and resonant scattering require an appropriate radiation-transfer calculation. The incident directions are determined by the same geometry used to calculate the photon density.

The scattering rate for an individual electron in the host-galaxy frame is  $\dot N_{\rm sc}=c\int d\epsilon\,d\Omega_{\rm i}\,n_{\rm ph}(\epsilon,\Omega_{\rm i})
(1-\beta_e\mu_{\rm i})\,\sigma_{\rm KN}(\epsilon_\star)$, where \(\sigma_{\rm KN}\) is the total Klein–Nishina cross-section evaluated at the incident photon energy in the electron rest frame. For a monochromatic, unidirectional seed component in the Thomson regime, this becomes $\dot N_{{\rm sc},j}\simeq c\sigma_{\rm T}n_j(1-\beta_e\mu_{{\rm i},j})$. This relation illustrates how the photon supply and incidence angle jointly control the scattering rate. It counts photons emitted into all outgoing directions; the directional emissivity of Section \ref{sec:ic_emissivity} is required to predict the observed flux.

The energy extracted from an electron must be calculated
from the energy gained by the photons, rather than their
final energy alone
\cite{Blumenthal1970,2010ApJS..189..286P}.  Using the redistribution kernel defined above, the net IC energy-loss rate for an electron with specified \(\gamma_e\) and \(\Omega_e\) is

\begin{eqnarray}
P_{\rm IC} = c\int d\epsilon\,d\Omega_{\rm i}\,
n_{\rm ph}(1-\beta_e\mu_{\rm i}) \int d\epsilon^{\rm IC}\,d\Omega_{\rm s}\, (\epsilon^{\rm IC}-\epsilon)\,\mathcal K.
\end{eqnarray}

For net IC cooling, $P_{\rm IC}>0$. A characteristic host-frame energy-loss time is then

\begin{eqnarray}
t_{\rm IC}
=
\frac{(\gamma_e-1)m_ec^2}{P_{\rm IC}}.
\end{eqnarray}

This time must be compared with the interaction time in the same frame, rather than directly with the observed duration. If radiative losses appreciably change the electron energy during the interaction, maintaining a prescribed constant Lorentz factor requires energy input during the interaction or a transport calculation that establishes the assumed distribution. Continuous injection alone does not prevent individual electrons from cooling as they cross the region.

The instantaneous electron kinetic energy within the emitting region is

\begin{eqnarray}
E_{e,\rm kin}(t)=
m_ec^2
\int d^3\boldsymbol r\,d\gamma_e\,d\Omega_e\,
(\gamma_e-1)f_e.
\end{eqnarray}

Integrating $P_{\rm IC}f_e$ over volume, electron phase space and host-frame time gives the total energy transferred from the electrons to the radiation. This transfer must be supplied by the initial particle energy, incoming particles or explicitly modelled energy input. If the electrons are dynamically coupled to ions or magnetic fields, the corresponding energy reservoir must also be included.

The observed isotropic-equivalent luminosity must not be identified with the angle-integrated radiated power of a directional source. The former is obtained from the observed flux, whereas the latter requires integration over all emission directions. A consistent evaluation of the observable luminosity and the actual energy requirement must use the same scattering geometry.

These quantities define common physical consistency tests for the seed candidates: the line luminosity and source–electron separation, the scattering electron content, associated radiation, radiative losses and photon escape. The calculations below establish kinematic estimates and restricted spectral tests, rather than a complete energy-budget solution for every candidate. In particular, the conditional molecular and UV components share an electron population, but their spectral normalization does not separately determine the absolute seed density and electron content.

\section{Seed-photon candidates}
\label{sec:seeds}

We apply the framework developed above to candidate transitions, distinguishing their kinematic suitability from the availability of the required local radiation field. For each candidate, we examine the electron energy, accompanying scattered emission, photon supply and radiation-transfer requirements. The numerical estimates below are illustrative configurations, rather than fits to the observed spectra.

Table~\ref{tab:seed_comparison} summarizes the seed energies, reference electron Lorentz factors and observational motivation for each candidate, distinguishing observations of GRB~221009A and its host from antecedents in other sources.

\begin{table}[tp]
\centering

\small
\setlength{\tabcolsep}{4pt}
\renewcommand{\arraystretch}{1.2}
\begin{adjustbox}{max width=\textwidth}
\begin{tabular}{@{}p{0.13\textwidth}
                  p{0.16\textwidth}
                  p{0.14\textwidth}
                  p{0.50\textwidth}@{}}
\hline
Seed candidate & Seed energy & $\gamma_e$ & Observational motivation \\
\hline

H$_2$ + UV &
$0.584$ eV;\newline
UV: $11.2$--$13.6$ eV &
$2.22\times10^3$\textsuperscript{a} &
JWST detected strong late-time H$_2$ emission at the projected GRB position;
the line ratios favour UV fluorescence \cite{2024NatAs...8..774B}.
The UV field at the scattering location is unmeasured. \\

Ly$\alpha$ &
$10.2$ eV &
$531$\textsuperscript{a} &
Indirectly motivated by host H$\alpha$ and Pa$\alpha$ detections
{\cite{2025A&A...701A.134M}}. Ly$\alpha$ is a hypothetical seed;
its local supply depends on production and resonant transfer. \\

H$\alpha$ &
$1.89$ eV &
$1.23\times10^3$\textsuperscript{a} &
H$\alpha$ and Pa$\alpha$ emission detected in the host galaxy
{\cite{2025A&A...701A.134M}}; these measurements do not establish
a local prompt seed field. \\

Mg XII &
$1.455$ keV &
$57.1$\textsuperscript{b} &
Marginal $3\sigma$ emission feature in the XMM-Newton afterglow
at approximately $2.3$ days, tentatively identified with Mg XII
\cite{2024ApJ...972...75C}; its availability during the prompt episode is unestablished. \\

Si XIV &
$2.01$ keV &
$50.8$\textsuperscript{b} &
Historical emission-line claim in GRB~011211 \cite{Reeves2002};
absorption identified in XB~1916$-$053 \cite{2006ApJ...647.1341I}.
A local seed field in GRB~221009A is unestablished. \\

S XVI &
$2.62$ keV &
$46.3$\textsuperscript{b} &
Historical emission-line claim in GRB~011211 \cite{Reeves2002};
absorption identified in XB~1916$-$053 \cite{2006ApJ...647.1341I}.
A local seed field in GRB~221009A is unestablished. \\

Fe K-shell &
$6.40$--$6.97$ keV &
$34.5$--$35.3$\textsuperscript{b} &
Iron-line claims in other GRBs, including GRB~991216 \cite{Piro2000},
and K-shell emission in the Galactic diffuse X-ray spectrum
\cite{2016PASJ...68...59Y}; no prompt seed source is established in this event. \\

\hline
\end{tabular}
\end{adjustbox}
\caption{Seed-photon candidates for an IC interpretation of the MeV feature in GRB~221009A.}
\label{tab:seed_comparison}

\medskip
\begin{minipage}{0.97\textwidth}
\footnotesize
\textit{Notes.}
Kinematic estimates assume a 10-MeV observed photon at $z=0.151$,
stationary host-frame seeds, head-on incidence and observer-aligned electrons;
$\gamma_e$ is defined in the host frame.\\
$^{a}$Thomson estimate.\\
$^{b}$Ultrarelativistic recoil-corrected estimate.
For H$_2$ + UV, the listed $\gamma_e$ refers to the $2.122\,\mu{\rm m}$ molecular line.
Historical GRB X-ray line claims have disputed significance \cite{Sako2005}.
The antecedents listed here do not establish the local prompt photon densities.
%The LAT consistency test concerns the associated IC--UV component of the H$_2$--UV scenario (Section~\ref{sec:h2_uv_calculation}).
\end{minipage}
\end{table}

\subsection{Molecular hydrogen and ultraviolet fluorescence}
\label{sec:molecular}

The JWST evidence summarized in Section~\ref{sec:observations} motivates the molecular seed candidate considered here.

This motivation is supported by earlier work on molecular gas in GRB environments. UV pumping, molecular excitation and the competing destruction of H$_2$ and dust were modelled by Draine~\cite{Draine2000} and Draine and Hao~\cite{DraineHao2002}. Vibrationally excited H$_2$ was detected in absorption towards GRB~080607 \cite{Sheffer2009}, while Wiersema et al.~\cite{Wiersema2018} searched for infrared H$_2$ emission in GRB hosts and reported a candidate $1-0\,S(3)$ line in the host of GRB~031203. These studies provide relevant excitation models and observational precedents; neither the absorption detection nor the host-emission measurements determine the prompt $2.122\,\mu{\rm m}$ photon density required here.

We adopt the $\mathrm{H}_2\;1-0\,S(1)$ transition at $\lambda_{\mathrm{H}_2}\simeq2.122\,\mu\mathrm{m}$, corresponding to $\epsilon_{\mathrm{H}_2}\simeq0.584$ eV. For head-on incidence, scattering to 10 MeV requires $\gamma_{e,\mathrm{H}_2}\simeq2.22\times10^3$.  In the head-on reference configuration, the incident photon energy in the electron rest frame is approximately $2.6$ keV, placing the molecular seed comfortably within the Thomson regime.

Fluorescence also motivates an associated UV field. Photons in the Lyman–Werner range, approximately $11.2\!-\!13.6$ eV, can excite molecular hydrogen and initiate radiative cascades \cite{1996ApJ...468..269D, 2015MNRAS.446..160A}. This range characterizes an excitation band rather than a single UV emission line. The incident UV spectrum must therefore be specified independently of the infrared (IR) line profile.  If these UV photons encounter the same electrons with the same incidence angles as the molecular photons, the Thomson energy mapping gives

\begin{eqnarray}
\epsilon_{\rm obs, UV}^{\rm IC}
\simeq
\epsilon_{\rm obs, H_2}^{\rm IC}
\frac{\epsilon_{\rm UV}}{\epsilon_{\rm  H_2}}
\simeq
(192\!-\!233)\,{\rm MeV}
\left(\frac{\epsilon_{\rm obs,H_2}^{\rm IC}}{10\,{\rm  MeV}}\right).
\end{eqnarray}

The Thomson estimate places the scattered UV radiation within the LAT band, but recoil lowers its characteristic energy. Evaluating equation~(\ref{eq:E_obs}) for head-on incidence and forward scattering, using the same electrons with $\gamma_e=\gamma_{e,H_2}\simeq2.22\times10^3$, gives $4\gamma_{e,H_2}\epsilon_{\rm UV}/(m_ec^2)\simeq0.19$--$0.24$ and characteristic scattered energies of approximately 160--190~MeV. These estimates apply to the specified geometry and do not identify an observed LAT spectral component.

A realistic molecular spectrum contains several transitions.
We therefore also tested molecular line spectra from
Draine and Bertoldi~\cite{1996ApJ...468..269D}, preserving their relative intensities
rather than assigning the entire molecular luminosity to the
$2.122\,\mu{\rm m}$ line.  The bounded test in Section~\ref{sec:h2_multiline} found accompanying scattered emission in excess of the adopted central spectral reference for every tested combination. The single-line calculation in Section~\ref{sec:h2_uv_calculation} must consequently be interpreted as a kinematic and spectral-moment illustration, not a successful description of the full molecular spectrum.

The fluorescence interpretation links the production of IR emission to UV illumination, but it does not guarantee that both fields illuminate the scattering electrons with comparable geometry. Absorption, molecular shielding and the spatial separation of the illuminating stars, molecular gas and outflow determine the local IR-to-UV radiation ratio. The UV field responsible for exciting the molecules cannot simply be equated with the field at the electron location.

The observational evidence for molecular emission motivates this candidate. Its main challenges are supplying sufficient local molecular photons and reproducing the MeV spectrum when accompanying transitions are included. Any UV radiation reaching the same electrons must also yield an IC contribution compatible with the LAT spectral reference. The restricted tests below expose difficulties with the molecular spectrum and photon supply.

%The principal attraction of this candidate is therefore a physically motivated route to related MeV and LAT components from a common electron population. Its main challenges are supplying sufficient molecular line photons, accounting for the full molecular spectrum and demonstrating that the required UV radiation reaches the same scattering region. The restricted tests below expose difficulties with both the accompanying molecular emission and the photon supply; they do not establish a joint spectral, temporal and energetic solution.

\subsection{Hydrogen  recombination lines: \texorpdfstring{$Ly\alpha$ and $H\alpha$}{Ly-alpha and H-alpha}}

Hydrogen provides two candidate transitions: $Ly\alpha$, with a rest-frame energy of approximately $10.2$ eV, and $H\alpha$, with a rest-frame energy of approximately $1.89$ eV. We initially assume that the emitting gas is approximately stationary in the host-galaxy frame. Any bulk motion of the seed source must subsequently be included when constructing the incident photon distribution.

For GRB~221009A, Malesani et al.~\cite{2025A&A...701A.134M} detected host-galaxy H$\alpha$ and Pa$\alpha$ emission, providing direct motivation for considering hydrogen recombination seeds. More generally, measurements of hydrogen and helium recombination lines in GRB hosts constrain extinction and line ratios \cite{Wiersema2011}. Such host measurements do not establish a compact source supplying the scattering electrons during the MeV episode, and their observed ratios cannot be identified directly with local seed-density ratios.

For electrons directed towards the observer in the Thomson regime, we define
$\mathcal{A}\equiv2(1-\beta_e\mu_i)$. Equation (\ref{eq:E_obs_s}) then becomes
$\epsilon_{\rm obs}^{\rm IC}\simeq\frac{\mathcal A\gamma_e^2\epsilon}{1+z}$.  For head-on incidence, $\mathcal A\simeq4$, and photons scattered to $10$ MeV require $\gamma_{e,\mathrm{Ly}\alpha}\simeq531$ and $\gamma_{e,\mathrm{H}\alpha}\simeq
1.23\times10^3$.  At these head-on reference values, the incident electron-rest-frame energies are approximately $10.8$ and $4.7$ keV, respectively. Thomson estimates therefore provide a useful starting point, although the full scattering kernel is retained in quantitative calculations.

The two transitions also imply a conditional spectral association. If both photon fields encounter the same electrons with the same angular distribution, their scattered energies approximately satisfy

\begin{eqnarray}
\frac{\epsilon_{\rm obs, Ly\alpha}^{\rm IC}}
     {\epsilon_{\rm obs, H\alpha}^{\rm IC}}
\simeq
\frac{10.2}{1.89}\simeq5.40.
\end{eqnarray}

Thus, identifying a 10-MeV feature with scattered $Ly\alpha$ places the associated $H\alpha$ component near 1.85 MeV.  Conversely, identifying it with scattered $H\alpha$ places the $Ly\alpha$ component near 54 MeV. These are characteristic energy mappings; the detectability and profiles of the accompanying components require their local photon densities and angular distributions.

For a MeV feature evolving between 6 and 12~MeV, the H$\alpha$-seed interpretation places the associated Ly$\alpha$ component at approximately 32--65~MeV in the Thomson approximation, under the same angular assumptions. This range overlaps the low-energy LAT band, particularly that addressed by the LAT Low Energy (LLE) analysis. Hydrogen seeds can therefore also produce accompanying emission that must remain compatible with contemporaneous observations. We do not perform a quantitative LLE comparison for this component here.

The production and transfer of the hydrogen radiation constrain its usefulness as a seed field. $Ly\alpha$ can be produced through recombination cascades and collisional excitation, but resonant transfer modifies its frequency and angular distribution before it reaches the scattering electrons \cite{2019SAAS...46....1D}. Dust absorption and gas structure can also alter its strength relative to the continuum; a clumpy medium does not generally guarantee preferential $Ly\alpha$ escape \cite{2014A&A...562A..52D}.

A self-consistent hydrogen-seed calculation must include the lines together with the radiation associated with their production mechanism. Where recombination dominates, the intrinsic line emission is constrained by the gas conditions and recombination cascade. The photon densities encountered by the electrons additionally depend on transfer and geometry, so an intrinsic line ratio cannot be directly imposed as a local seed-density ratio.

Hydrogen seeds offer identifiable companion components and relatively modest electron-rest-frame seed energies in the reference configurations. Their principal unresolved requirement is a sufficiently luminous, suitably directed radiation field at the scattering location during the MeV episode. This requirement must be tested alongside the predicted emission in the GBM and LAT bands.

%\subsection{Highly ionized metal lines: the Mg XII candidate}
\subsection{Mg XII: a highly ionized metal-line candidate}
\label{sec:li-Mag}

A possible metal-line seed is motivated by the XMM-Newton observations of GRB 221009A obtained approximately 2.3 days after the prompt emission. \cite{2024ApJ...972...75C} reported a candidate feature at a host-frame energy of $1.455^{+0.006}_{-0.014}$ keV (90 per cent confidence for one parameter of interest), consistent with slightly redshifted Mg~XII $K\alpha$ emission.  Its significance is marginal, approximately $3\sigma$. The authors proposed reflection of late-time central-engine radiation within the funnel as a possible origin. This provides a motivation for investigating metal-line production, but does not demonstrate the presence of the required photons during the prompt MeV episode.

We adopt $\epsilon_{\rm Mg}=1.455$~keV as an illustrative host-frame seed energy. For head-on incidence and scattering towards an observer aligned with the electron velocity, neglecting recoil would give $\gamma_e\simeq44.5$ for a 10-MeV photon. However, recoil is appreciable and must be retained:

\begin{eqnarray}
(1+z)\epsilon^{\rm IC}_{\rm obs}
=
\frac{4\gamma_e^2\epsilon_{\rm Mg}}
{1+4\gamma_e\epsilon_{\rm Mg}/(m_ec^2)}.
\end{eqnarray}

Solving this expression for an observed photon energy of
10~MeV gives $\gamma_e\simeq57.1$.
The corresponding incident photon energy in the electron
rest frame is approximately 166~keV, with
$4\gamma_e\epsilon_{\rm Mg}/(m_ec^2)\simeq0.65$,
confirming that the Thomson approximation is insufficient.

%Writing $Y\equiv(1+z)\epsilon_{\rm obs}^{\rm IC}$, the corresponding solution is $\gamma_e=\frac{1}{2}\left[\frac{Y}{m_ec^2}+\sqrt{\left(\frac{Y}{m_ec^2}\right)^2+ \frac{Y}{\epsilon_{\rm Mg}}}\right]$.  For $\epsilon_{\rm obs}^{\rm IC}=10\,{\rm MeV}$ and $z=0.151$, this yields \(\gamma_e\simeq57.1\). The incident electron-rest-frame energy is approximately \(166\) keV, and the recoil parameter is $\frac{4\gamma_e\epsilon_{\rm Mg}}{m_ec^2} \simeq0.65$.   A Thomson-only treatment is therefore insufficient for a quantitative comparison.

For this reference electron population, the limiting scattered-photon energy is approximately $\gamma_e m_ec^2/(1+z)\simeq25$~MeV. These electrons therefore cannot produce emission above 100~MeV by scattering soft photons. This restriction does not exclude Mg~XII seeds as an explanation of the MeV
feature alone.

The principal motivation for Mg~XII is its tentative observational association with this event. However, the reported feature is marginal and was observed long after the MeV episode. A viable seed model must establish a sufficient local photon density during that episode and include the associated continuum, illumination geometry and radiation transfer. These requirements remain
unestablished.

\subsection{Silicon, Sulphur and Iron K-shell lines}

We extend the comparison to Si XIV and S XVI Ly$\alpha$
transitions, with representative energies of approximately
2.01 and 2.62 keV. These ions have been identified through
X-ray absorption features in other astrophysical sources,
such as XB 1916$-$053 \cite{2006ApJ...647.1341I}.
We also consider iron K-shell emission from low-ionization
Fe K$\alpha$ near 6.4 keV, the Fe XXV He$\alpha$ complex
near 6.7 keV and Fe XXVI Ly$\alpha$ near 6.97 keV;
these emission components have been studied separately in
the Galactic diffuse X-ray spectrum \cite{2016PASJ...68...59Y}.
We treat these transitions as hypothetical seed fields
for the prompt episode, rather than as established
detections in GRB 221009A.

For a stationary seed source, head-on incidence and electrons directed towards the observer, we use the recoil-corrected expression in Section \ref{sec:li-Mag} to estimate the Lorentz factor required to produce a 10-MeV photon. The resulting values are summarized in Table \ref{tab:seed_comparison}. The limiting scattered-photon energy for each reference population is approximately $\gamma_e m_ec^2/(1+z)$.

Higher seed energies reduce the electron Lorentz factor required in this geometry, but increase the importance of recoil. In particular, the iron candidates cannot be evaluated using the Thomson energy scaling alone. Their spectra and scattering rates require the full Klein–Nishina kernel.

The same calculation illustrates a limitation of their simplest connection to LAT. The characteristic recoil limits of these reference electron populations remain below 100 MeV. Producing LAT photons by IC scattering of soft radiation therefore requires higher-energy electrons, either within an extended distribution or in a different angular configuration that retains the metal-seed component near 10 MeV. The tabulated values do not exclude such configurations, but show that the MeV and LAT components are not connected automatically by selecting a higher-energy seed.

The production of the seed radiation imposes separate constraints. Earlier GRB reprocessing models explicitly examined the illuminating radiation, gas mass and geometry required to generate X-ray lines \cite{ReesMeszaros2000,KumarNarayan2003,Bottcher2004}. These studies concern line production, not an established IC origin of the MeV feature. Low-ionization iron fluorescence and highly ionized recombination or collisionally excited lines require different gas conditions. Their luminosities depend on the elemental abundances, ionization balance, illuminating spectrum and column density. A candidate model must generate the line together with the associated continuum and other transitions, rather than specify an isolated monochromatic photon field.

For these X-ray seeds, recoil can appreciably reduce the contribution of a small fractional seed-energy spread to the scattered-energy spread, as described by equation~(\ref{eq:seed_energy_response}). This effect does not guarantee a narrow observable feature: the angular distributions, electron-energy dispersion, relative line strengths and temporal integration must still be included.

These metal transitions broaden the comparison of seed candidates, but a smaller required Lorentz factor is not by itself evidence of greater physical viability. Their assessment depends on whether an appropriate reprocessing region can supply the photons during the MeV episode while producing a spectrum and energy budget compatible with the simultaneous GBM and LAT observations.

\section{Comparison and discussion}
\label{sec:discussion}

The candidate seed fields differ in their characteristic
energies, observational support and photon-production
requirements. We assess their kinematic accessibility,
spectral compatibility and physical photon supply.
Section~\ref{sec:seeds} provides reference estimates of the
electron energies required to reach the MeV band.
The molecular multi-line test in
Section~\ref{sec:h2_multiline} supplies our most direct
spectral constraint: at the adopted normalization,
the accompanying transitions produce emission exceeding
the central spectral reference in every tested configuration.
Any predicted high-energy contribution must also remain
compatible with contemporaneous LAT observations, without
requiring the model to explain the observed LAT emission.
The molecular mismatch remains conditional on the tested
configurations and adopted spectral references.

%The candidate seed fields differ in their characteristic energies, observational support and photon-production requirements. Our comparison distinguishes three questions: whether a transition can be scattered into the MeV band, whether the resulting spectrum can reproduce the observed feature, and whether the same physical configuration can account for the contemporaneous LAT emission. The estimates in Section \ref{sec:seeds} address primarily the first question. The molecular multi-line test in Section~\ref{sec:h2_multiline} supplies our most direct spectral constraint on the explored MeV interpretation: at the adopted normalization, the molecular contribution alone exceeds the central total spectral reference. This test does not require an associated IC component to be identified in LAT. The constraints on a joint MeV--LAT origin are considered separately and do not establish a preferred model through a joint spectral fit.

%\subsection{Kinematic comparison and a possible MeV--LAT connection}
\subsection{Kinematic requirements and accompanying emission}

Matching the MeV centroid alone does not uniquely identify the seed photons. Lower-energy seeds require larger electron Lorentz factors in the reference geometry, whereas X-ray seeds require lower Lorentz factors but experience stronger recoil. The angular distribution introduces an additional degeneracy: different combinations of electron energy and incidence angle can produce the same observed photon energy.

For the head-on, monoenergetic reference configurations summarized in Table~\ref{tab:seed_comparison}, the metal-line seeds require electron Lorentz factors of approximately 35--57 to produce 10-MeV photons. Their limiting scattered energies are approximately 15--25~MeV, preventing these electrons from producing photons above 100~MeV by scattering soft radiation. This does not exclude those configurations as contributors to the MeV feature alone.

Hydrogen and molecular seeds require larger electron Lorentz factors in the reference geometry and can produce higher-energy companion emission if suitable additional seed photons reach the same electrons. For example, an H$\alpha$-seed interpretation of a 10-MeV feature places the associated Ly$\alpha$ component near 54~MeV under common Thomson scattering conditions. Its strength depends on the local photon fields and transfer; we do not perform a quantitative LLE comparison here.

The molecular-fluorescence configuration offers a more direct conditional connection. If the $2.122\,\mu{\rm m}$ $H_2$ transition produces the MeV feature, UV photons in the Lyman–Werner band can be scattered into the LAT range by the same electrons.  For the head-on, monoenergetic reference configuration of Section~\ref{sec:molecular}, a molecular component near 10 MeV corresponds to characteristic UV-scattering energies around 160--190 MeV after recoil is included.   This is an energy mapping, not a prediction of a narrow LAT line or a demonstrated match to an observed excess. Its spectral shape and normalization depend on the UV spectrum and the relative illumination geometries.

The observational motivation also differs among candidates. The molecular emission reported by \cite{2024NatAs...8..774B} identifies line-emitting gas at the projected burst position, whereas the Mg XII feature reported by \cite{2024ApJ...972...75C} is marginal and was observed in the late X-ray afterglow. Neither observation directly measures the prompt seed field at the scattering location. Consequently, observational association and kinematic suitability must be assessed separately.

Within these reference configurations, $H_2$ fluorescence supplies a conditional common-electron energy mapping, while the X-ray candidates reveal stronger kinematic restrictions.  However, Section~\ref{sec:precursor_pumping} does not establish a source configuration that supplies the required molecular photon density, while
the bounded test in Section~\ref{sec:h2_multiline} yields scattered emission exceeding the adopted central spectral references in every
tested configuration. These conditional results prevent the
favourable energy mapping from establishing a preferred
physical model; they do not constitute a statistical
exclusion of molecular seeds.

\subsection{Seed-photon supply and relative IC fluxes}

The required seed-photon density is a decisive constraint on the candidate models. For prescribed spectral shapes, geometry and electron energies, the optically thin emissivity scales with the product of the electron and seed-photon densities. Consequently, matching the MeV luminosity alone does not determine either density uniquely. Increasing the electron content can reduce the required seed density, but also changes the particle energy budget and scattering optical depth. Increasing the seed density enhances the radiative losses and may invalidate a constant-energy electron distribution.

Observed line emission motivates a seed species but does not by itself close this local budget. The line luminosity intercepted by the scattering region depends on its distance, solid angle, transfer and timing. The energetic cost of producing line radiation from an illuminating continuum was already central to GRB X-ray reprocessing studies \cite{Ghisellini2002,KumarNarayan2003}; the numerical constraints from those environments cannot simply be transferred to molecular gas.

The relative component strengths provide a conditional consistency test for emission produced by a common electron population. For a common observing interval, we define $\mathcal{R}_{\rm obs}=F_{\rm LAT}/F_{\rm line}$,
where $F_{\rm LAT}$ is the total energy flux integrated over a specified LAT band and $F_{\rm line}$ is the energy flux attributed to the MeV feature above the fitted continuum. This ratio provides an observational reference for the predicted IC contribution, without identifying the total LAT flux with that contribution. Statistical constraints would require propagating the flux and continuum-decomposition uncertainties, which is not done here.

For a uniform, optically thin scattering region, let $u_j$
denote the local host-frame seed-photon energy density of
component $j$, and let $\mathcal{P}_{j,B}$ denote the predicted
observed energy flux in band $B$, per scattering electron
and per unit $u_j$. This coefficient is calculated from the directional scattering kernel, the normalized electron and seed-photon distributions,
the source geometry, cosmological redshift, luminosity distance and attenuation.  The component flux is $F_{j,B}=\mathcal{N}_e u_j\mathcal{P}_{j,B}$, where $\mathcal{N}_e$ is the total number of scattering electrons.
For two components produced by the same electrons,

\begin{eqnarray}
\frac{F_{2,B_2}}{F_{1,B_1}} = \frac{u_2}{u_1} \frac{\mathcal P_{2,B_2}}{\mathcal P_{1,B_1}}.
\end{eqnarray}

The common electron normalization cancels, allowing the flux ratio to constrain the relative seed energy densities once the remaining model parameters are specified. For spatially varying or evolving fields, the full integrations of Section \ref{ref:scattering} must replace this factorized expression.

In the Thomson regime, for the same electron population, identical seed angular distributions, negligible attenuation and integration over the complete scattered components, the luminosity ratio approximately follows the seed
energy-density ratio. This simplification is generally insufficient for the present comparison: GBM and LAT cover finite energy bands, the seed fields may illuminate the electrons differently, and recoil modifies both the scattering probability and the energy redistribution. Because an associated LAT component is not established, the total LAT flux cannot be assigned to this process as a measured normalization.

These requirements prevent a definitive ranking of seed candidates
based solely on their characteristic scattered energies.
Each candidate must satisfy the absolute photon-supply and
electron-energy budgets, relative component strengths and
escape conditions. None of the candidates is demonstrated
here to satisfy all of these requirements.

\subsection{Spectral width and temporal evolution}

The observed width and temporal evolution provide constraints beyond the characteristic energy and luminosity. A seed transition can satisfy the energy mapping while producing a scattered component that is too broad, or evolves differently from the MeV feature. These properties must therefore be calculated jointly for the proposed electron distribution and source geometry.

In the aligned Thomson approximation,
$\epsilon_{\rm obs}^{\rm IC}\simeq\mathcal{A}\gamma_e^2\epsilon/(1+z)$, so small variations satisfy $\mathrm{d}\ln\epsilon_{\rm obs}^{\rm IC}
=2\,\mathrm{d}\ln\gamma_e+\mathrm{d}\ln\epsilon +\mathrm{d}\ln\mathcal{A}$.
For narrow, statistically independent distributions,
this gives the approximate width relation

\begin{eqnarray}
\left(\frac{\sigma_{\epsilon^{\rm IC}_{\rm obs}}}{\epsilon^{\rm IC}_{\rm obs}}\right)^2
\simeq
4\left(\frac{\sigma_{\gamma_e}}{\gamma_e}\right)^2
+
\left(\frac{\sigma_\epsilon}{\epsilon}\right)^2
+
\left(\frac{\sigma_{\mathcal{A}}}{\mathcal{A}}\right)^2.
\end{eqnarray}

This expression is a local diagnostic rather than a substitute for the scattering calculation. The relevant distributions are weighted by their contribution to the observed photons; correlations introduce covariance terms, while broader distributions and recoil require the full kernel.

The importance of recoil differs among the seed candidates. Using the respective head-on reference Lorentz factors adopted in Section~\ref{sec:seeds}, we obtain $b\simeq0.010$, $0.018$, $0.042$ and $0.65$ for the H$_2$ reference line, H$\alpha$, Ly$\alpha$ and Mg~XII,
respectively. Equation~(\ref{eq:seed_energy_response}) then gives response factors $1/(1+b)\simeq0.990$, $0.982$, $0.959$ and $0.606$. Thus, the fractional seed-energy spread is transmitted almost unchanged for the molecular and hydrogen reference lines, whereas recoil appreciably reduces this contribution for Mg~XII. These factors quantify only the response to small seed-energy variations at fixed electron energy and scattering geometry; the total observed width also depends on the electron distribution, angular spread and temporal integration.

For an illustrative relative width $\sigma_{\epsilon_{\rm obs}^{\rm IC}}/\epsilon_{\rm obs}^{\rm IC}\simeq0.1$, the electron contribution is bounded by $\sigma_{\gamma_e}/\gamma_e\lesssim0.05$
under these assumptions. This example is not the adopted relative width in either of the two observing intervals analysed in Section~\ref{sec:h2_uv_calculation}. Seed and geometrical broadening reduce the width available to the electron distribution. Thus, the monoenergetic limit is useful for assessing whether a candidate fails even before finite electron dispersion is introduced.

Temporal integration adds another contribution.  For the intrinsic scattered line component, let \(W(t_{\rm obs})\) be its normalized photon-flux weight within an observing interval, such that \(\int dt_{\rm obs}\,W(t_{\rm obs})=1\). Its time-integrated variance obeys

\begin{eqnarray}
\sigma_{\rm int}^{2}
=
\int dt_{\rm obs}\,W(t_{\rm obs})
\,\sigma_{\rm inst}^{2}(t_{\rm obs})
+
\int dt_{\rm obs}\,W(t_{\rm obs})
\left[\epsilon^{\rm IC}_c(t_{\rm obs})-\overline \epsilon^{\rm IC}_c\right]^2,
\end{eqnarray}

where $\sigma_{\rm inst}(t_{\rm obs})$ and
$\epsilon_c^{\rm IC}(t_{\rm obs})$ are the instantaneous
intrinsic width and centroid of the scattered line,
respectively, and
$\overline{\epsilon_c^{\rm IC}}
=\allowbreak\int \mathrm{d}t_{\rm obs}\,W(t_{\rm obs})
\epsilon_c^{\rm IC}(t_{\rm obs})$
is the photon-weighted mean centroid over the observing
interval.  The second term represents broadening caused by centroid evolution. This moment relation applies before instrumental convolution and need not equal the width returned by a Gaussian fit to a non-Gaussian profile.

An observational benchmark for this temporal test is the approximately inverse-time evolution reported for the MeV line. Analysing the published line measurements, Zhang et al.~\cite{Zhang2024JetPhysics} obtain $E_{\rm line}\propto(t_{\rm obs}-t_\star)^{-k}$, with $k=1.05^{+0.22}_{-0.17}$ and $t_\star=226^{+8}_{-10}$~s, where $t_{\rm obs}$ is measured relative to the burst reference time. This empirical relation provides a temporal constraint on proposed emission mechanisms. Its applicability to the spectral decomposition and observing intervals adopted here must nevertheless be assessed rather than assumed.

For a fixed seed spectrum, illumination geometry and electron-distribution shape in the Thomson regime, the characteristic scattered energy scales approximately as the square of the characteristic electron Lorentz factor. Reproducing a centroid evolution $E_{\rm line}\propto(t_{\rm obs}-t_\star)^{-k}$
would therefore require $\gamma_e\propto(t_{\rm obs}-t_\star)^{-k/2}$,
approximately $(t_{\rm obs}-t_\star)^{-1/2}$ for $k\simeq1$, under these assumptions. This scaling specifies a required evolution rather than identifying the mechanism that produces it. Translating it into a local cooling or acceleration law also requires the arrival-time relation discussed in Section~\ref{sec:Obs_flux_evol}. Evolution of the electron dispersion, illumination angles or recoil corrections can modify this correspondence. In particular, the independently calibrated distributions considered below do not establish such a dynamical evolution.

A common-electron interpretation also predicts conditional relations between the temporal evolution of different scattered components. If two fixed seed spectra share the same angular distribution and remain in the Thomson regime, their characteristic energy ratio stays approximately constant as the electron energy changes. Departures from this behaviour can arise from recoil,
temporal changes in the relative illumination geometries, or evolution of the seed spectra.

Using the head-on ultrarelativistic scattering relation
introduced above, the response to a change in electron
Lorentz factor at fixed seed energy and geometry is

\begin{eqnarray}
\frac{d\ln \epsilon^{\rm IC}_{\rm obs}}{d\ln\gamma_e}
=
\frac{2+b}{1+b}.
\end{eqnarray}

The response approaches two in the Thomson limit ($b\ll1$) and tends to unity in the strong-recoil limit ($b\gg1$). For molecular seeds and their associated UV radiation, the UV-scattered component can therefore evolve differently from the MeV component even when both originate from the same electrons. This relation describes characteristic scattering energies; the evolution of observable spectral peaks must be obtained from the integrated spectra.

The temporal test must consequently use a specified evolution of the electron population and seed geometry to predict both components over matched observing intervals. Independently adjusting their energies and normalizations in each interval would not establish a physical association.  A model proposing a common-electron origin should account
for the measured MeV centroid, width and flux, while keeping
its predicted associated high-energy component compatible
with the contemporaneous LAT spectrum and its uncertainties.

\subsection{The \texorpdfstring{H$_2$}{H2}--UV inverse-Compton scenario}
\label{H2-UV_scenario}

For molecular fluorescence, the IR and UV normalizations must be linked through the source model. In a steady-state approximation, we write

\begin{eqnarray}
L_{\rm H_2,ref}^{\rm esc}
=
\eta_{\rm ref}L_{\rm UV}^{\rm abs}\,,
\end{eqnarray}

where $L^{\rm esc}_{{\rm H}_2,\rm ref}$ is the luminosity escaping from the molecular gas in the reference H$_2$ $1-0\,S(1)$ transition at $2.122\,\mu{\rm m}$, $L_{\rm UV}^{\rm abs}$ is the UV luminosity absorbed by the molecular gas during fluorescent excitation, and $0\leq\eta_{\rm ref}\leq1$ is the fraction of that absorbed power re-emitted and escaping in the reference line. This efficiency includes the cascade yield and escape of the IR photons. It is not the fraction emerging in all molecular transitions, and it cannot be chosen independently of the excitation and transfer calculation. For transient excitation, the time-dependent molecular level populations must instead be followed, as discussed in Section~\ref{sec:precursor_pumping}.

Moreover, the absorbed UV luminosity at the molecular gas does not determine the UV density at the electron location. The positions of the illuminating source, molecular material and scattering region must jointly specify the two seed fields. A successful configuration must supply the molecular photons required by GBM while keeping any accompanying UV-scattered contribution compatible with LAT.

\subsubsection{Illustrative inverse-Compton spectra}
\label{sec:h2_uv_calculation}

We now extend the characteristic-energy comparison to an
illustrative calculation of the scattered spectral profiles in
two observing intervals. We consider
$T_0+[290.6,300.6]$~s and $T_0+[300.6,325.6]$~s, using the
centroids, widths and photon fluxes of the Gaussian MeV component
reported by Axelsson et al.~\cite{2025ApJS..277...24A}.  We adopt the central Gaussian parameters from their table~9. Their observational uncertainties are not propagated through this illustrative calibration. The purpose is to derive the
associated UV-seed contribution after calibrating the
electron population against the MeV feature. This calculation
does not constitute a joint fit to the GBM and LAT count spectra.  Here, $E$ denotes photon energy in the observer frame, expressed in MeV; for the calculated IC components, $E=\epsilon_{\rm obs}^{\rm IC}$. Seed-photon energies continue to be denoted by $\epsilon$ and are defined in the host-galaxy frame.

Unlike the head-on reference configurations discussed above,
we adopt lateral illumination: the incident seed photons
propagate perpendicular to the electron velocity, while the
observer lies along the electron direction of motion.
Both seed fields share this angular configuration.
Electron Lorentz factors and seed-photon energies are defined
in the host-galaxy frame, and the scattered photon energies
include the cosmological redshift, $z=0.151$.
The electron distribution is specified by 
\begin{eqnarray}\nonumber
 \frac{dN_e}{d\ln\gamma_e}=\frac{N_e}{\sqrt{2\pi}\,s_\gamma}
\exp\left[-\frac{  \left(\ln\gamma_e-\ln\gamma_0\right)^2}{2s_\gamma^2}
\right]\,,   
\end{eqnarray}

where $\gamma_0$ is the median Lorentz factor and
$s_\gamma$ is the standard deviation in $\ln\gamma_e$.

The molecular seed field is represented by the
$2.122\,\mu{\rm m}$ transition, with
$\epsilon_{\rm H_2}=0.584$~eV and negligible intrinsic width
for this illustrative calculation.
We include recoil and the differential Klein--Nishina
cross-section when constructing the observer-directed
scattered spectrum.
Within each interval, $\gamma_0$ and $s_\gamma$ are chosen
to reproduce the photon-weighted centroid and standard
deviation of the published Gaussian component.
The molecular IC normalization is then set by its inferred
energy flux.
Matching these moments does not enforce a Gaussian scattered
profile, and differences between the predicted and reference
profiles remain possible, particularly in their wings.  Each interval is represented by a fixed electron distribution and illumination geometry. The calibrated electron dispersion is therefore an effective parameter of the time-integrated spectrum and need not equal the instantaneous dispersion.

For the associated UV field, we assume uniform
seed energy density per unit photon energy between
11.2 and 13.6~eV in the host frame.
This is a prescribed test spectrum rather than a measurement
of the local UV radiation.
Keeping the electron distribution fixed by the molecular
component yields UV-scattered photon centroids
of approximately 210 and 161~MeV in the first and second
intervals, respectively.
The corresponding standard deviations are approximately
35 and 55~MeV.
Table~\ref{tab:h2_uv_intervals} summarizes the adopted
MeV parameters and the resulting electron and UV-component
parameters.

\begin{table}[tp]
\centering

\begin{adjustbox}{max width=\textwidth}
\begin{tabular}{lcc}
\hline
Quantity
& $290.6$--$300.6$~s
& $300.6$--$325.6$~s \\
\hline
\multicolumn{3}{l}{\textit{Adopted MeV reference parameters}} \\
Photon centroid, $E_{\rm c,H_2}$ [MeV]
& 11.5 & 8.8 \\
Photon standard deviation, $\sigma_{E,{\rm H_2}}$ [MeV]
& 2.0 & 3.2 \\
Line energy flux,
$F_{\rm line}$ [$10^{-6}$ erg cm$^{-2}$ s$^{-1}$]
& 3.65 & 3.58 \\
\hline
\multicolumn{3}{l}{\textit{Calibrated electron distribution}} \\
Median Lorentz factor, $\gamma_0$
& $3.30\times10^3$ & $2.69\times10^3$ \\
Logarithmic dispersion, $s_\gamma$
& 0.087 & 0.177 \\
\hline
\multicolumn{3}{l}{\textit{Predicted UV-scattered component}} \\
Photon centroid, $E_{\rm c,UV}$ [MeV]
& 210 & 161 \\
Photon standard deviation, $\sigma_{E,{\rm UV}}$ [MeV]
& 35 & 55 \\
Illustrative seed-density ratios, $q$
& $0.03,\ 0.05,\ 0.10$
& $0.03,\ 0.05,\ 0.10$ \\
\hline
\end{tabular}
\end{adjustbox}
\caption{Illustrative IC parameters for lateral illumination
by an isolated $2.122\,\mu{\rm m}$ H$_2$ line and a UV field
with uniform energy density per unit photon energy over
11.2--13.6~eV. The MeV inputs are central Gaussian parameters
from Axelsson et al.~\cite{2025ApJS..277...24A}; the electron calibration
is described in Section~\ref{sec:h2_uv_calculation}.}
\label{tab:h2_uv_intervals}

\smallskip
\begin{minipage}{0.97\textwidth}
\footnotesize
\textit{Notes.}
Times are relative to $T_0$; times and scattered photon
energies are in the observer frame.
Electron Lorentz factors and seed energies are in the host frame.
The observer is aligned with the electron velocity.
Spectral widths are photon standard deviations, not centroid
uncertainties.
The local ratio $q=u_{\rm UV}/u_{\rm H_2}$ uses the integrated
UV-band energy density and only the selected H$_2$ line.
The displayed $q$ values are illustrative.
\end{minipage}
\end{table}

\begin{figure}[tp]
\centering
\includegraphics[width=\textwidth]{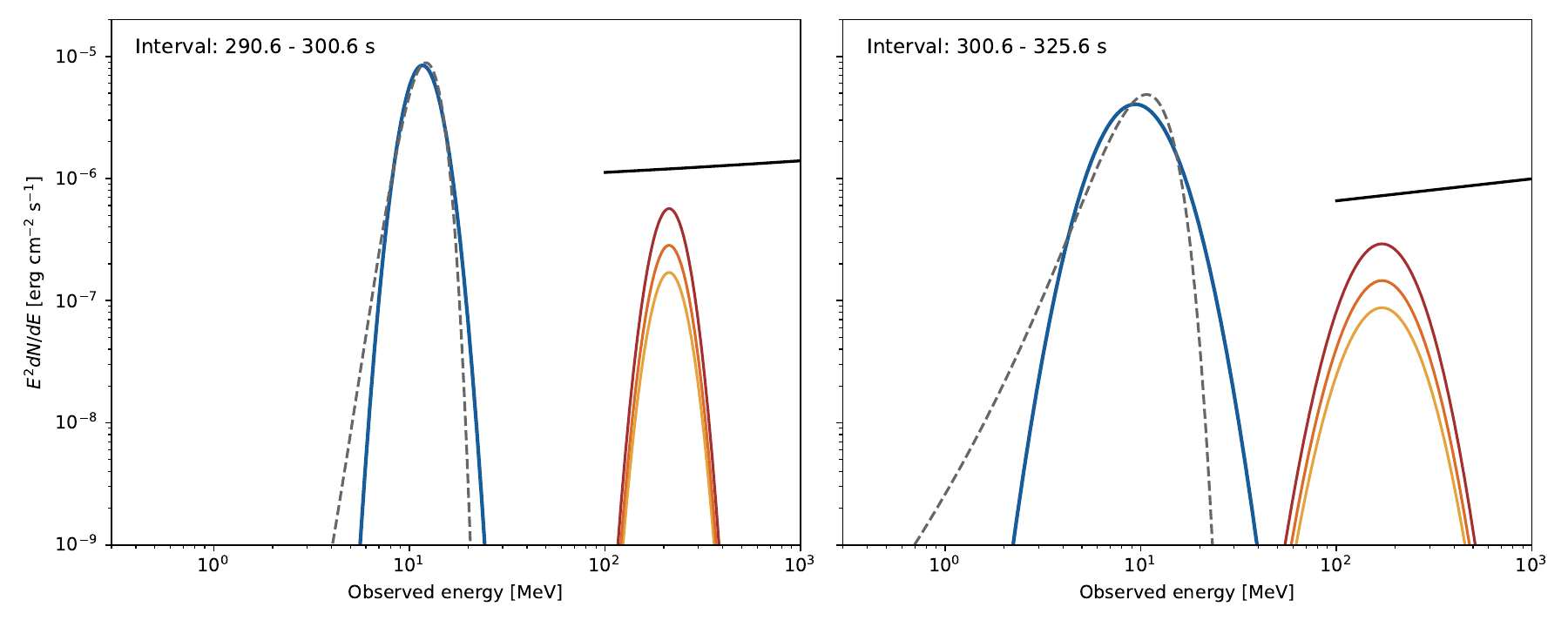}
\caption{Illustrative IC spectra for GRB~221009A in the indicated observer-time intervals relative to $T_0$. Blue curves show IC emission from the isolated $2.122\,\mu{\rm m}$ H$_2$ line; grey dashed curves show the published Gaussian MeV reference used for calibration. Gold, orange and red curves show IC--UV emission for $q=u_{\rm UV}/u_{\rm H_2}=0.03$, 0.05 and 0.10, respectively. Black curves are LAT power-law references reconstructed
from Axelsson et al.~\cite{2025ApJS..277...24A}. Model parameters and calibration are summarized in Table~\ref{tab:h2_uv_intervals}; its photon centroids need
not coincide with the SED maxima. The full GBM continuum and accompanying molecular transitions are omitted.}
 \label{fig:ic_components}
\end{figure}

We describe the relative seed normalization through $q \equiv \frac{u_{\rm UV}}{u_{\rm H_2}}$,  where both energy densities are evaluated locally at the
scattering region in the host-galaxy frame.
Here, $u_{\rm H_2}$ refers to the selected molecular line,
whereas $u_{\rm UV}$ is integrated over the adopted
UV band.
For the prescribed geometry and spectral shapes, the
optically thin UV-scattered flux scales linearly with $q$.
We display $q=0.03$, 0.05 and 0.10 as illustrative
normalizations in Figure~\ref{fig:ic_components}.
These values are not inferred seed-density ratios and are
not yet derived from a precursor or fluorescence model.
The lateral direction prescribed for the UV field is likewise an
independent assumption; it is not established by the central-precursor
geometry discussed in Section~\ref{sec:precursor_pumping}.

The reconstructed LAT power-law spectra provide a reference
against which to compare these components.
We use the central power-law parameters from Table~3 of
Axelsson et al.~\cite{2025ApJS..277...24A}.  These published fits combine LLE data with standard LAT data.  For
$T_0+[290.6,295.6]$~s and $T_0+[295.6,300.6]$~s, the adopted
energy fluxes are $9.0\times10^{-6}$ and $1.4\times10^{-5}$
erg~cm$^{-2}$~s$^{-1}$, with photon exponents $-2.0$ and $-1.8$,
respectively. For $T_0+[300.6,325.6]$~s, they are
$9.0\times10^{-6}$ erg~cm$^{-2}$~s$^{-1}$ and $-1.82$. For each published LAT time interval $i$, we reconstruct
a differential photon flux $N_i(E)\propto E^{\alpha_i}$, in ${\rm ph\,cm^{-2}\,s^{-1}\,MeV^{-1}}$, and normalize it such that $c_{\rm MeV}\int_{E_a}^{E_b} E\,N_i(E)\,{\rm d}E=F_i$, where $c_{\rm MeV}\simeq 1.6\times10^{-6}\,{\rm erg\,MeV^{-1}}$
converts MeV to erg, and $E_a=100\,{\rm MeV}$ and
$E_b=10^5\,{\rm MeV}$ define the normalization band.  For the first ten-second interval, the reference is the duration-weighted average of the two spectra,
$\overline{N}(E)=[5\,{\rm s}\,N_1(E)+5\,{\rm s}\,N_2(E)]/(10\,{\rm s})$.
This is not an average of their photon exponents or an
exposure-weighted fit. The 100~MeV--100~GeV normalization band
is distinct from the 100--1000~MeV segment displayed in Figure~\ref{fig:ic_components}.
The displayed LAT curves are reference models, not
independent spectral measurements, and should not be added
unchanged to the IC contribution and interpreted as a
joint best fit.
No associated IC--UV component is established by this comparison.
The archived calculation uses published power-law parameters,
without an event-level LAT likelihood, exposure or response
calculation. Its agreement or disagreement with those reference
curves therefore cannot be converted into a detection significance
or a confidence limit on $q$. Such a limit would require the
continuum and IC components to be fitted together with the
appropriate instrumental response and verified good-time
coverage, particularly near the beginning of the first interval.

The independently calibrated electron distributions suggest a
possible physical interpretation of the spectral softening.
For the same seed transition and fixed lateral illumination,
the median host-frame Lorentz factor decreases from
$\gamma_0\simeq3.30\times10^3$ to $\simeq2.69\times10^3$, a reduction of approximately
19 per cent. Net radiative cooling could qualitatively produce
such a decrease \cite{Blumenthal1970}. Alternatively, successive
electron populations could be injected at lower characteristic
energies, or the balance between acceleration and losses could
change during the emission episode. These possibilities are
not distinguished by the two independent calibrations, which
do not establish that the same electrons evolve between the
inferred distributions.

The width provides an additional constraint: the calibrated
logarithmic dispersion increases from $s_\gamma\simeq0.087$
to $\simeq0.177$. For a narrow population subject to a common,
continuous Thomson-loss history, radiative cooling contracts
the instantaneous fractional energy spread rather than
broadening it. Thus, this simple cooling picture would not,
by itself, explain both calibrated changes if they represented
instantaneous distributions. However, our parameters describe
time-integrated spectra. Centroid evolution within each
observing interval can contribute to the effective width,
as can spatially varying cooling histories, continued injection
or acceleration. The increase in the calibrated dispersion
therefore does not independently establish intrinsic
broadening of the electron population.

A physical connection between the two intervals would require evolving the electron distribution and local radiation field, with the appropriate photon arrival-time mapping, and predicting the centroid, width and flux together. We have not performed that calculation. The present results therefore provide a conditional single-line MeV--LAT prediction, without establishing an associated LAT component or resolving the photon-supply difficulty and multi-line spectral mismatch.
The precursor considered below may prepare the molecular excitation before scattering, but it does not yet determine the electron evolution or the local seed-density ratio.

Appendix~\ref{app:four_intervals} extends this moment calibration to four observing intervals, using a consistent positive-energy convention and documenting its relation to the original two-interval calculation. The central calibrated values do not form a monotonically decreasing sequence. A bounded sensitivity test using the published marginal parameter ranges contains decreasing examples, but these combinations cannot be assigned a joint confidence level without the parameter covariances. The extension therefore describes additional interval-integrated spectra without establishing the evolution of a single electron population.

\FloatBarrier
\subsubsection{A precursor-assisted molecular seed scenario}
\label{sec:precursor_pumping}

The precursor of GRB~221009A provides a possible temporal
framework for preparing the molecular seed field before
the scattering episode.
Stern and Tkachev~\cite{2023JETPL.118..553S} report a gamma-ray precursor
approximately three minutes before the main emission.
This observation establishes an earlier radiative episode,
but does not measure the UV radiation incident
on nearby molecular material.
We therefore consider precursor-driven UV pumping as
a hypothesis for the origin of the seed photons,
rather than an observed property of the local environment.

In this picture, a UV pulse excites molecular
hydrogen, and the subsequent radiative cascade populates
rovibrational levels that can continue emitting after
the illuminating pulse has passed. GRB-driven pumping and
time-dependent molecular destruction have been studied by
Draine~\cite{Draine2000} and Draine and Hao~\cite{DraineHao2002}, but those calculations do not
establish the precursor UV spectrum or geometry assumed here.
The stored quantity is molecular excitation, rather than
a trapped population of IR photons.
If the scattering electrons encounter the resulting
IR radiation after the direct UV pulse has declined,
the local ratio $u_{\rm UV}/u_{\rm H_2}$ can be lower
than during illumination.
This temporal separation could help reduce the associated
UV-scattered component without eliminating the molecular
seed field.

The timing must satisfy the propagation geometry.
Let a precursor photon be emitted at host-frame time
$t_{\rm p}$ from the origin, excite molecular material
at position $\boldsymbol{r}_{\rm m}$, and produce an
IR photon after a delay $\tau$.
For that photon to reach a scattering event at
$(\boldsymbol{r}_{\rm e},t_{\rm e})$, its trajectory
must satisfy
\begin{eqnarray}
t_{\rm e}
=
t_{\rm p}
+\frac{|\boldsymbol{r}_{\rm m}|}{c}
+\tau
+\frac{
|\boldsymbol{r}_{\rm e}-\boldsymbol{r}_{\rm m}|
}{c}.
\label{eq:precursor_causality}
\end{eqnarray}
The relevant IR emission time is
$t_{\rm IR}=t_{\rm e}-|\boldsymbol{r}_{\rm e}-\boldsymbol{r}_{\rm m}|/c
=t_{\rm p}+|\boldsymbol{r}_{\rm m}|/c+\tau$.
Thus, the upper-level population supplying photons to that
scattering event must be evaluated at $t_{\rm IR}$.
Consequently, the observed precursor--main-emission
separation cannot be identified directly with the local
molecular storage time.
In particular, a short precursor does not automatically
illuminate material located far to the side of the jet
early enough to supply laterally incident seed photons.
Equation~(\ref{eq:precursor_causality}) must be checked
for any realization of the geometry adopted in
Section~\ref{sec:h2_uv_calculation}.   Here $\tau$ is measured in the host-galaxy frame. The photon arrival-time relation additionally requires
$\tau\leq\Delta t_{\rm obs}/(1+z)$, where $\Delta t_{\rm obs}$ is measured from the direct arrival of the illuminating precursor pulse.

After the pumping and rapid cascade have ceased,
a minimal loss model for the population $N_u$ of the
upper level of the selected molecular transition is
\begin{eqnarray}
\frac{dN_u}{dt}=-\left(A_{\rm tot}+C_u+D_u\right)N_u,
\label{eq:upper_level_decay}
\end{eqnarray}

where $A_{\rm tot}$ is the total spontaneous decay rate
out of level $u$, $C_u=\sum_a n_a k_{ua}$ is the collisional depopulation
rate, and $D_u$ is the rate of additional destructive losses. Here, $n_a$ is the number density of collision partner species $a$, and $k_{ua}$ is its total collisional depopulation rate coefficient for level $u$.  This expression neglects subsequent feeding from other levels and assumes that the rates remain constant.  The escaping luminosity in the reference line is
$L_{\mathrm{H}_2,\mathrm{ref}}^{\rm esc}(t)
=f_{\rm esc}A_{u\ell}\epsilon_{\mathrm{H}_2}N_u(t)$,
where $f_{\rm esc}$ is the escape fraction of the reference-line
photons and $A_{u\ell}$ is the Einstein coefficient of the
selected transition.  Its inverse is not the lifetime of the upper level,
because other decay and collisional channels contribute.
The competition between pumping, radiative cascades
and collisional processes is central to molecular
fluorescence calculations
\cite{1996ApJ...468..269D}.

For illustration, retaining at least half of the post-pumping
population over a local interval $\Delta t_{\rm loc}$ requires
$A_{\rm tot}+C_u+D_u\lesssim\allowbreak
2.3\times10^{-3}(\Delta t_{\rm loc}/300\,{\rm s})^{-1}
\,{\rm s}^{-1}$.  The factor of $300\,{\rm s}$ is only a normalization scale
for this illustrative condition; the physical delay must
satisfy the propagation and arrival-time constraints above.  Moreover, preserving an excited population is only
one requirement: the IR luminosity and its geometrical dilution must still supply the seed density needed by the IC calculation.

To illustrate the pumping-energy scale, we adopt a hypothetical, time-integrated, isotropic-equivalent precursor photon spectrum in the host frame,
$dN_\gamma/d\epsilon\propto\epsilon^{-2.75}$, between 11.2 eV and 10 keV.   Normalizing it to an isotropic-equivalent Lyman--Werner energy of $E_{\rm LW,iso}=2.3\times10^{51}$ erg in the 11.2--13.6 eV band gives approximately $1.7\times10^{52}$ erg between 11.2 eV and 10 keV.    These energies specify an assumed low-energy precursor component; they are not measurements of its UV output. A turnover below the adopted lower boundary is also
needed if this steep spectrum is extended further. A large isotropic-equivalent burst energy alone does not establish either this spectral component or its allocation to the precursor.

The intercepted photon budget provides a separate
constraint.
For a molecular structure of projected area
$A_{\rm proj}$ at distance $r$ from a point-like
illuminating source, a single-pulse estimate gives
\begin{eqnarray}
N_u(t_{\rm IR})
\lesssim
p_u
\frac{A_{\rm proj}}{4\pi r^2}
\frac{E_{\rm LW,iso}}
{\langle\epsilon_{\rm LW}\rangle},
\label{eq:pumping_photon_budget}
\end{eqnarray}
where the isotropic-equivalent energy is evaluated
in the direction of the molecular material, and
$\langle\epsilon_{\rm LW}\rangle$ is the photon-number-weighted
mean energy in the 11.2--13.6 eV band in the host frame.  This bound assumes at most one useful upper-level excitation per intercepted UV photon, so $0\leq p_u\leq1$. The effective yield $p_u$ includes transmission to the molecules, absorption, cascade feeding and survival until IR emission at $t_{\rm IR}$. Overlapping structures must not count the same intercepted photons repeatedly. This estimate shows why sufficient global energy does not guarantee a sufficient local molecular seed field: the intercepted fraction and useful excitation yield can both be small.

The present calculation does not establish a source configuration
that supplies the required molecular photon density while satisfying
the timing and angular constraints.
Prescribing lateral incidence in the scattering kernel does not
itself specify a viable seed source. This unresolved source requirement
does not exclude every possible molecular geometry.

An intervening neutral layer could reduce the ionizing
radiation reaching the molecules, but its effect on
the useful pumping spectrum must be evaluated together
with dust attenuation and molecular self-shielding.
Protection against ionization cannot be treated as
transparent transmission of the entire pumping band.
The subsequent main emission must also be included
when assessing molecular survival.

Complete removal of the UV field before scattering would suppress
the associated IC--UV component. Residual illumination, partial temporal
overlap or an additional UV source would also have to supply the required
incident directions. Direct photons from the central source propagate
approximately along the electron motion on the axis. For exactly
copropagating photons and an on-axis observer,
$\mu_i=\mu_s=1$ and $\psi=0$, equation~(\ref{eq:E_obs}) gives
$\epsilon_{\rm obs}^{\rm IC}=\epsilon_{\rm UV}/(1+z)$,
without the amplification of the lateral case. The sub-GeV prediction
therefore requires UV redirected into lateral directions or supplied
by another suitably located source. Both $q$ and the lateral UV
illumination in Section~\ref{sec:h2_uv_calculation} remain prescribed
inputs, rather than outputs of the central-precursor model.
Precursor-assisted excitation can separate pumping and
scattering in principle, but a common dynamical and radiative
geometry satisfying the photon-supply requirements remains
unestablished. We retain it as a conditional mechanism,
not as a solution to the absolute molecular seed budget.

An irradiated surface is a possible geometrical analogy. In the Fe-line model of Rees and M{\'e}sz{\'a}ros~\cite{ReesMeszaros2000}, line emission arises in a thin photoionized layer on the wall of a funnel, allowing escape towards the cavity. This separates line production in dense material from propagation through a less opaque region. A molecular analogue would additionally require H$_2$ survival, a sufficient escaping infrared luminosity, appropriate illumination of the electrons and the observed timing. The hot, ionized Fe-emitting layer is not itself an H$_2$ source. We do not calculate the thermal, chemical or radiative-transfer structure of a molecular layer here. The following estimates quantify the geometrical sensitivity of its photon budget without establishing a physical source or resolving the spectral discrepancies of the tested molecular templates.

\noindent\textit{Illustrative photon-budget estimates.}
We considered a stationary scattering zone at axial distance $R$ from the
central source, surrounded by a cylindrical molecular wall of radius $d$
and half-height $H$. A continuously replenished, uniform electron cylinder
has radius $a$ and length $\ell$, with its axis and electron velocities
aligned with the observer. All lengths are measured in the host frame.
The reference dimensions were $(d,H,a,\ell)=(10^{12},1.7\times10^{11},
10^{11},10^{11})$~cm. We varied $R$ and applied scale factors $s_\perp$
to $(d,H,a)$ and $s_\parallel$ to $\ell$, retaining the hypothetical
precursor energy specified above. These are prescribed examples, not
geometries inferred from the observations.

For normalization, we used the Rh3o configuration described in
Section~\ref{sec:h2_multiline}, with $\gamma_e=3.38\times10^3$,
$\tau_{\rm ref}=0.5$ and prescribed fractional width $0.12$ in
$T_0+[290.6,300.6]$~s. Its archived energy-flux coefficient is
$C_{\rm mol}=5.13\times10^{-7}\,{\rm erg\,cm^{-2}\,s^{-1}}$;
this is not the total MeV-component flux. Let $u_{\rm ref,0}$ and
$u_{\rm ref}=e^{-\tau_{\rm ref}}u_{\rm ref,0}$ denote the reference-line
energy densities before and after the external screen. With
$d_L=730$~Mpc as a reference luminosity distance, the leading
ultrarelativistic, on-axis lateral-scattering normalization gives
$N_e u_{\rm ref,0}\simeq4\pi d_L^2 C_{\rm mol}/
(12\sigma_{\rm T}c\gamma_e^4)=1.05\times10^{48}\,{\rm erg\,cm^{-3}}$.
The recoil weights remain in the molecular spectral sum; the reference-line
weight is approximately $0.98$. This normalization applies to a stationary
zone, without an additional moving-source arrival-time compression.
Imposing a transverse Thomson depth
$\tau_{\rm T}\simeq2N_e\sigma_{\rm T}/(\pi a\ell)\leq0.1$
sets $N_{e,\max}=0.1\pi a\ell/(2\sigma_{\rm T})$ and a corresponding
minimum required local density $u_{\rm ref,req}$. Only this photon budget
was varied; the finite-geometry spectral profiles were not recalculated.

For $R\gg d,H$, central illumination strikes the wall at grazing incidence,
giving an intercepted fraction $f_{\rm cov}\simeq d^2H/R^3$.
To construct an especially favourable upper bound, we allowed all
intercepted Lyman--Werner energy to be stored in the reference upper level,
whose excitation energy is $\mathcal{E}_u\simeq0.60$~eV. This gives
$N_u\leq f_{\rm cov}E_{\rm LW,iso}/\mathcal{E}_u$, a relaxed energy bound
that does not impose the single-absorption yield in
equation~(\ref{eq:pumping_photon_budget}). With isotropic spontaneous
emission, unit internal escape and the same external screen as above,
the local density is bounded by
\begin{eqnarray}
u_{\rm ref,max}\lesssim
\frac{e^{-\tau_{\rm ref}}A_{u\ell}\epsilon_{\mathrm{H}_2}}
{4\pi c d^2}
\frac{f_{\rm cov}E_{\rm LW,iso}}{\mathcal{E}_u},
\label{eq:h2_seed_energy_bound}
\end{eqnarray}
where $A_{u\ell}\simeq3.5\times10^{-7}\,{\rm s^{-1}}$ for the
$1-0\,S(1)$ transition, consistent with the tabulated line intensity
and apparent upper-level column of Draine and Bertoldi~\cite{1996ApJ...468..269D}.
Equation~(\ref{eq:h2_seed_energy_bound}) is a maximum-storage estimate,
not a calculated fluorescent luminosity.

Table~\ref{tab:h2_seed_budget} compares this bound with the required
field. Improved interception and moderate size increases reduced the
minimum deficit $\mathcal{D}=u_{\rm ref,req}/u_{\rm ref,max}$ from
approximately $1.3\times10^6$ to $8.5$ or $4.3$ in the compact examples.
The latter contain $N_e\simeq4.7\times10^{45}$ and
$9.4\times10^{45}$ electrons, respectively, with actual electron kinetic
energies of approximately $1.3\times10^{43}$ and $2.6\times10^{43}$~erg
for the assumed aligned population. These are not isotropic-equivalent
jet energies. Neither compact example supplies the required local
reference-line density even under the ideal energy-storage and internal
escape assumptions. If $\eta_u$ is the fraction of intercepted UV energy
stored in the reference upper level at the relevant time, multiplied by
the internal line escape fraction, the deficit increases to
$\mathcal{D}/\eta_u$. The external screen is already included in
$\mathcal{D}$. This transient efficiency is distinct from the steady-state
power conversion efficiency $\eta_{\rm ref}$ and is not measured.

\begin{table}[tp]
\centering

\small
\setlength{\tabcolsep}{9pt}
\renewcommand{\arraystretch}{1.15}
\begin{adjustbox}{max width=\textwidth}
\begin{tabular}{@{}cccccc@{}}
\hline
$R$ & $s_\perp$ & $s_\parallel$ &
$u_{\rm ref,req}$ & $u_{\rm ref,max}$ & $\mathcal{D}$ \\
{[cm]} & & & [${\rm erg\,cm^{-3}}$] & [${\rm erg\,cm^{-3}}$] & \\
\hline
$10^{16}$ & 1 & 1 & $2.7\times10^2$ & $2.1\times10^{-4}$ & $1.3\times10^6$ \\
$10^{15}$ & 1 & 1 & $2.7\times10^2$ & $0.21$ & $1.3\times10^3$ \\
$3\times10^{14}$ & 2 & 1 & $1.3\times10^2$ & $16$ & $8.5$ \\
$3\times10^{14}$ & 2 & 2 & $67$ & $16$ & $4.3$ \\
\hline
\end{tabular}
\end{adjustbox}
\caption{Conditional reference-line photon budget for
$T_0+[290.6,300.6]$~s.}
\label{tab:h2_seed_budget}
\smallskip
\begin{minipage}{0.97\textwidth}
\footnotesize
\textit{Notes.} The scale factors multiply the reference dimensions
$(d,H,a)$ and $\ell$, respectively, defined in
Section~\ref{sec:precursor_pumping}.
Both local fields include the external attenuation $e^{-0.5}$.
The available field is the relaxed energy-storage upper bound in
equation~(\ref{eq:h2_seed_energy_bound}), not a predicted fluorescent
luminosity. None of the examples closes that ideal budget.
Under the single-use photon assumption with $p_u=1$ and unit internal
escape, each deficit increases by a factor of approximately $20.5$.
The deficit ratios were evaluated before rounding.
\end{minipage}
\end{table}

The single-use photon budget is more restrictive. The adopted spectrum
gives $\langle\epsilon_{\rm LW}\rangle\simeq12.3$~eV and
$\langle\epsilon_{\rm LW}\rangle/\mathcal E_u\simeq20.5$.
Under equation~(\ref{eq:pumping_photon_budget}), the efficiency obeys
$\eta_u\leq p_u\mathcal E_u/\langle\epsilon_{\rm LW}\rangle
\simeq0.049p_u$, before any further reduction by internal escape.
Even for $p_u=1$ and unit internal escape, the compact deficits are then
approximately $1.7\times10^2$ and $87$, respectively. A value such as
$\eta_u=0.1$ is therefore not attainable within this single-use assumption;
it would require an additional energy-redistribution or photon-reuse
mechanism that is not calculated here. The relaxed energy ceiling and
the single-use photon bound describe different assumptions, not two
predictions of a calculated fluorescent source.

For an illustrative observed precursor--IC separation of $300$~s,
the central wall element in both compact examples requires an excitation
age of approximately $194$~s, from
$\Delta t_{\rm obs}/(1+z)=[\sqrt{R^2+d^2}+d-R]/c+\tau$.
Their central angular spread remains approximately
$H/[\sqrt{3}\sqrt{d^2+H^2}]\simeq0.097$ in incident-direction cosine.
The axial light-travel spread is approximately $3.8$ or $7.7$~s observed;
the latter is comparable to the first $10$-s analysis window.
These estimates do not replace a convolution of the injection history,
angular kernel and propagation delays.

The required population also constrains the material supplying the
line. Let $\mathcal N_u$ denote the column of molecules in the reference
upper level through the wall. Neglecting all
attenuation gives the optimistic lower bounds
$N_u\simeq(3.1\mbox{--}6.2)\times10^{56}$,
$\mathcal N_u\simeq(3.6\mbox{--}7.3)\times10^{31}\,{\rm cm^{-2}}$
and $M_{\rm H_2,min}\simeq0.52\mbox{--}1.0\,M_\odot$ if every
molecule occupies that level. These are requirements for the assumed
luminosity, not populations produced by the precursor calculation.
If $f_u$ is the fraction of H$_2$ in that level and
$f_{\rm esc,int}$ is the internal reference-line escape fraction
(the $f_{\rm esc}$ of the loss model above), restoring the external
screen gives $M_{\rm H_2,req}\gtrsim
(0.86\mbox{--}1.7)\,M_\odot/(f_u f_{\rm esc,int})$.
For a material thickness $\Delta r$ along the column direction, even
the lower bounds without attenuation require
$n_{\rm H_2}\gtrsim(1.8\mbox{--}3.6)\times10^{19}
[\Delta r/(2\times10^{12}\,{\rm cm})]^{-1}f_u^{-1}\,{\rm cm^{-3}}$.
The thickness is an illustrative conversion scale, not an inferred or
resolved wall depth. At fixed area and luminosity, reducing it increases
the density without reducing the required emitting column.

At the local excitation age of $194$~s, equation~(\ref{eq:upper_level_decay})
requires $A_{\rm tot}+C_u+D_u\lesssim3.6\times10^{-3}\,{\rm s^{-1}}$
for half of the initial population to survive without further feeding.
Taking a collider density equal to the optimistic minimum upper-level
density at $\Delta r=2\times10^{12}$~cm would require an effective
collisional depopulation coefficient
$k_{\rm eff}\lesssim(1\mbox{--}2)\times10^{-22}\,{\rm cm^3\,s^{-1}}$.
This is a necessary condition, not a calculated molecular rate or
survival fraction. Evaluating it requires the gas temperature,
collider abundances and state-dependent coefficients. Molecular
survival against photodissociation and ionization is a separate
requirement depending on the incident radiation and shielding.

Internal line opacity requires both upper- and lower-level populations.
For a Doppler profile with local velocity parameter $b_D$, the central
optical depth, including stimulated emission, is
\begin{eqnarray*}
\tau_0=\frac{A_{u\ell}\lambda^3}{8\pi^{3/2}b_D}
\left(\frac{g_u}{g_\ell}\mathcal N_\ell-\mathcal N_u\right),
\end{eqnarray*}
where $\mathcal N_\ell$ is the lower-level column and
$g_u/g_\ell=7/3$ for $1-0\,S(1)$. At
$b_D=3\,{\rm km\,s^{-1}}$, the coefficient multiplying the column
difference is approximately $2.5\times10^{-25}\,{\rm cm^2}$.
As a separate, non-inverted population example, setting
$\mathcal N_\ell=\mathcal N_u$ at the required column scale gives
$\tau_0\simeq(1.2\mbox{--}2.4)\times10^7$.
This example is not the all-upper-level population used to derive the
minimum mass; the latter would be inverted and could not be treated
as ordinary positive absorption. Neither the level populations nor
the local velocity profile are calculated here, so this estimate is
conditional and does not determine an escape fraction. Reabsorption,
conversion to other transitions, collisional losses and internal dust
absorption must be distinguished from the prescribed external screen
$\tau_{\rm ref}=0.5$.

\noindent\textit{Electron cooling in the required radiation field.}
The accompanying transitions also increase the electron losses.
Keeping the Rh3o intensity ratios, common geometrical dilution and the
external attenuation law of Section~\ref{sec:h2_multiline}, the reference
line plus just the ten rotational lines $0-0\,S(0)$ through $S(9)$ give
$u_{\rm mol}/u_{\rm ref}\geq
1+\sum_{j=0}^{9}\mathcal I_{0-0\,S(j)}
\exp[\tau_{\rm ref}-\tau(\lambda_{0-0\,S(j)})]\simeq14.4$.
Here $u_{\rm mol}$ is the local energy density summed over molecular
lines; this inequality does not include UV photons.
For lateral incidence in the Thomson limit,
$P_{\rm IC}\simeq\sigma_{\rm T}c\gamma_e^2u_{\rm mol}$.
Using the required reference-line densities in the two compact examples,
this partial molecular field gives initial host-frame cooling times of
approximately $6.3$ and $13$~s, compared with crossing times
$t_{\rm cross}\simeq\ell/c=3.3$ and $6.7$~s, respectively.
For a constant field and no acceleration, the ultrarelativistic solution
$\gamma_e(t)=\gamma_{e,0}/[1+t/t_{{\rm IC},0}]$ then gives an approximately
$35$~per~cent decrease during the crossing in either example.
Additional transitions could increase the losses. These are consistency
estimates evaluated at the required field, not a prediction that the
precursor supplies that field. Continuous injection alone does not
maintain the prescribed monoenergetic population throughout the volume;
energy input during transit or a transport calculation is required.

The improved interception therefore reduced the ideal deficit without
closing either the photon budget or the physical source model.
The stationary Rh3o template, whose hydrogen-nuclei density is
$n_{\rm H}=10^6\,{\rm cm^{-3}}$, supplies spectral ratios, not the
populations, efficiency or escape of the compact gas considered here.
The spectral comparisons and these source requirements are conditional
tests, not a single self-consistent molecular configuration.

\subsubsection{Spectral consistency test of molecular IC emission}
\label{sec:h2_multiline}

To test whether the single-line approximation survives the
inclusion of accompanying transitions, we used four published
molecular emission spectra from the stationary photodissociation-region
calculations of Draine and Bertoldi~\cite{1996ApJ...468..269D}: Lw3o, Mw3o, Qw3o
and Rh3o.\footnote{The tabulated spectra and model definitions
are available at \url{https://www.astro.princeton.edu/~draine/pdr.html}.}
Their prescribed densities, radiation-field strengths and
temperature parameters are listed in Table~\ref{tab:h2_multiline}.
We included all tabulated transitions, including pure rotational
lines; the source tables omit transitions weaker than $0.001$
times the reference $1$--$0\,S(1)$ intensity. This test uses
existing spectra, not a new calculation of the thermal or
chemical state of the gas or a fit to the JWST line ratios.
It does not evaluate a time-dependent molecular spectrum
produced by impulsive precursor excitation.

\begin{table}[tp]
\centering

\small
\begin{adjustbox}{max width=\textwidth}
\begin{tabular}{lrrrrrrr}
\hline
Model & $n_{\rm H}$ [cm$^{-3}$] & $\chi$ & $T_0$ [K] &
$R_{\max}^{\rm (0)}$ & $\min R_{\max}$ &
$\tau_{\rm ref}^{\rm best}$ & $(\sigma_j/E_j)^{\rm best}$ \\
\hline
Lw3o & $10^5$ & $10^3$ & 500  & 7.13 & 3.48 & 0.5 & 0.12 \\
Mw3o & $10^5$ & $10^4$ & 500  & 7.20 & 2.00 & 1.0 & 0.12 \\
Qw3o & $10^6$ & $10^4$ & 500  & 4.93 & 2.11 & 0.5 & 0.12 \\
Rh3o & $10^6$ & $10^5$ & 1000 & 3.68 & 1.82 & 0.5 & 0.12 \\
\hline
\end{tabular}
\end{adjustbox}
\caption{Bounded molecular IC test for $T_0+[290.6,300.6]$~s.
$R_{\max}$ is the maximum molecular-to-reference energy-flux-density
ratio over 0.1--100~MeV, after normalization to the Gaussian-component
flux in 9.5--13.5~MeV. The central SBPL+Gaussian+SBPL reference
uses pivots of 100~keV and 1~GeV.
Minima are taken over ten attenuation--width configurations
per template.}
\label{tab:h2_multiline}
\smallskip
\begin{minipage}{0.97\textwidth}
\footnotesize
\textit{Notes.}
Templates are from Draine and Bertoldi~\cite{1996ApJ...468..269D}.
$\chi$ is the illuminating intensity at 1000~\AA\
relative to the Habing reference; $T_0$ is the template
temperature parameter, unrelated to the GRB trigger.
All cases use host-frame $\gamma_e=3.38\times10^3$ and lateral incidence.
$R_{\max}^{(0)}$ denotes $R_{\max}$ evaluated at $\tau_{\rm ref}=0$ and $\sigma_j/E_j=0.06$.  Profiles and attenuation are prescribed
(Section~\ref{sec:h2_multiline}). The ratios are conditional diagnostics, not statistical exclusions.
\end{minipage}
\end{table}

This calculation is separate from the lognormal-electron
calibration of Section~\ref{sec:h2_uv_calculation}.
We fixed the host-frame electron Lorentz factor at
$\gamma_e=3.38\times10^3$, adopted lateral incidence and observer-aligned
scattering, and specialized equation~(\ref{eq:E_obs}) to this geometry
in the ultrarelativistic limit, retaining recoil:
\begin{eqnarray}
E_j=\frac{2\gamma_e^2\epsilon_j}
{(1+z)[1+2\gamma_e\epsilon_j/(m_ec^2)]}.
\label{eq:h2_multiline_energy}
\end{eqnarray}
Here $\epsilon_j$ is the seed-photon energy in the host-galaxy
frame. For the adopted geometry and redshift, $\gamma_e=3.38\times10^3$
maps the $1-0\,S(1)$ transition at $2.12183\,\mu{\rm m}$
to approximately $11.5$ MeV, the adopted centroid in
$T_0+[290.6,300.6]$ s.  Relative scattered powers were weighted by
$r_j^3(r_j+r_j^{-1})/2$, where
$r_j=[1+2\gamma_e\epsilon_j/(m_ec^2)]^{-1}$, corresponding
to the ultrarelativistic fixed-angle Klein--Nishina
approximation per unit seed energy density.
The prescribed Gaussian is a profile of energy-flux density,
$F_E=dF/dE$, rather than of differential photon flux. Explicitly,
\begin{eqnarray}\nonumber
G_j(E)=\frac{\exp[-(E-E_j)^2/(2\sigma_j^2)]}
{\sqrt{2\pi}\,\sigma_j},\qquad
F_{E,{\rm mol}}(E)=C_{\rm mol}\sum_j
\mathcal{I}_j e^{-\tau(\lambda_j)}
\frac{r_j^3(r_j+r_j^{-1})}{2}\,G_j(E),
\end{eqnarray}
with $\sigma_j/E_j=0.06$ or $0.12$, and energies in MeV.
We use the tabulated energy-intensity ratio
$\mathcal{I}_j=I_j/I_{1-0\,S(1)}$ from the
\texttt{I/I(10S1)} column; the sixth whitespace-separated field
of the archived line tables. These ratios are taken as relative seed energy densities under
a common geometrical dilution assumption; no extra seed-photon
energy factor is applied. $G_j$ is the unit-area Gaussian on
the real energy axis used in the archived calculation.
These widths are prescribed smoothing parameters, not a
calculated angular distribution or an instrumental response.
They must not be confused with the kernel-derived profiles
shown in Figure~\ref{fig:ic_components}. The coefficient $C_{\rm mol}$ is an overall energy-flux
normalization, with units of ${\rm erg\,cm^{-2}\,s^{-1}}$;
it is distinct from the unit-conversion factor $c_{\rm MeV}$.

We also applied a hypothetical, purely absorbing screen between the molecular emitter and the scattering region, attenuating the seed photons before inverse-Compton scattering. Its optical depth is
$\tau(\lambda)=\tau_{\rm ref} (2.122\,\mu{\rm m}/\lambda)^{1.7}$,
where $\tau_{\rm ref}=0,\,0.5,\,1,\,1.5,\,2$. This hypothetical attenuation law changes the escaping line ratios but does not alter the molecular level populations or represent an increased internal shielding
column. The four spectra, five optical depths and two widths give 40 combinations. Individual line intensities were not adjusted independently.

For $T_0+[290.6,300.6]$~s, each scattered molecular spectrum
was normalized to the energy flux of the published Gaussian
component, fixing $C_{\rm mol}$ through
$\int_{E_i}^{E_f} F_{E,{\rm mol}}(E)\,{\rm d}E
\simeq2.49\times10^{-6}\,{\rm erg\,cm^{-2}\,s^{-1}}$,
where $E_i=9.5\,{\rm MeV}$ and $E_f=13.5\,{\rm MeV}$
define the normalization window.
The molecular Gaussian window integrals are evaluated analytically,
using the unrounded target flux computed from the archived
parameters of the observed Gaussian line. The normalization $C_{\rm mol}$ is determined separately for each template, width and attenuation combination. Once fixed, it is retained in all comparisons with alternative reference spectra.  This normalization applies to the sum of all molecular
components within the selected window, rather than
to the scattered $2.122\,\mu{\rm m}$ line alone.  This procedure tests spectral compatibility at the adopted normalization; it does not establish that the required absolute seed-photon supply is physically available.    We then evaluated $R_{\max}=\max_E[F_{E,\rm mol}(E)/F_{E,\rm ref}(E)]$ on 1500 logarithmically spaced energy points between 0.1 and 100 MeV, where $F_{E,\rm ref}$ is the reconstructed central SBPL+Gaussian+SBPL spectral model of
Axelsson et al.~\cite{2025ApJS..277...24A}.\footnote{SBPL denotes a smoothly broken power law, which connects two power-law segments smoothly around
a break energy. See the \texttt{SmoothlyBrokenPowerLaw}
documentation and source commit \texttt{ac599aa}.
These references do not establish the software version
or configuration used in the published analysis.} This comparison asks whether the molecular contribution alone
exceeds the entire reference spectrum at any sampled energy.
The baseline reconstruction adopted pivots of 100~keV and
$10^6$~keV for the first and second SBPL components, respectively,
and a break scale of 0.5. The note accompanying the published
parameters specifies a different first pivot; below we test the
alternative interpretation and compare the reconstruction with
the displayed continuum.
No continuum refit, instrumental convolution or covariance
propagation was performed.

None of the 40 combinations satisfied this baseline central
spectral envelope at all sampled energies.
For $\sigma_j/E_j=0.06$, the unattenuated spectra reached
maximum model-to-reference ratios of approximately
3.68--7.20. Attenuation and broader prescribed profiles reduced
the mismatch, but the best case, Rh3o with
$\tau_{\rm ref}=0.5$ and $\sigma_j/E_j=0.12$, still reached
a ratio of approximately 1.82 at 8.74~MeV.
Table~\ref{tab:h2_multiline} summarizes the results. We then checked the sensitivity to the first-SBPL normalization.
The note to table~8 of Axelsson et al.~\cite{2025ApJS..277...24A} specifies a
pivot of 1~keV, whereas the public \texttt{astromodels}
documentation and source inspected here give a default of
100~keV. We repeated the same 40 configurations with a 1-keV pivot,
retaining the tabulated numerical normalization
$K_1=0.319\,{\rm ph\,cm^{-2}\,s^{-1}\,keV^{-1}}$ and all
other central parameters. The molecular spectra, their
normalization in the 9.5--13.5~MeV window and the evaluation
grid were unchanged.

This check compares two interpretations of the tabulated
normalization rather than two parametrizations of a fixed
spectrum. A consistent change of pivot requires
$K_{\rm new}=N_{\rm old}(E_{\rm piv,new})$, where
$E_{\rm piv,new}$ is the new pivot energy,
$N_{\rm old}(E)$ is the differential photon flux of the
first SBPL component before reparametrization, and
$K_{\rm new}$ is its normalization at the new pivot.
This transformation preserves the physical spectrum
and its integrated flux.  With the numerical $K_1$ held fixed, however, the first SBPL is lower by a factor $2.33\times10^3$ for the 1-keV interpretation. The second SBPL and the Gaussian remain unchanged, so the total reference spectrum decreases at every sampled energy and every
molecular-to-reference ratio increases. None of the 40 cases
satisfies the central envelope under either interpretation.
The minimum maximum ratio is 1.82 for the 100-keV reconstruction
and approximately $5.77\times10^2$ for the literal 1-keV
reconstruction. These minima refer to the best configuration
within each reconstruction, which need not be the same.
Table~\ref{tab:h2_pivot_sensitivity} reports the minimum for
each molecular template. The large ratios in the 1-keV column
characterize that alternative reconstruction, not a statistical
significance of disagreement with the data.

\begin{table}[tp]
\centering

\small
\begin{adjustbox}{max width=\textwidth}
\begin{tabular}{lcc}
\hline
Template & $E_{{\rm piv},1}=100$~keV & $E_{{\rm piv},1}=1$~keV \\
\hline
Lw3o & 3.48 & $3.55\times10^3$ \\
Mw3o & 2.00 & $5.77\times10^2$ \\
Qw3o & 2.11 & $9.02\times10^2$ \\
Rh3o & 1.82 & $7.77\times10^2$ \\
\hline
\end{tabular}
\end{adjustbox}
\caption{Sensitivity of the minimum $R_{\max}$ to the first-SBPL
pivot for the configurations in Table~\ref{tab:h2_multiline},
at fixed $K_1=0.319\,{\rm ph\,cm^{-2}\,s^{-1}\,keV^{-1}}$
and all other central parameters.
Each column is minimized separately over ten configurations
per template; molecular normalizations are unchanged.
These are alternative reconstructions, not a consistent
reparametrization or statistical significances.}
\label{tab:h2_pivot_sensitivity}
\end{table}

As an independent check of the continuum normalization, we
extracted the central-model vector trace for
$T_0+[290.6,300.6]$~s, the second interval in the sequence
shown in figure~8 of Axelsson et al.~\cite{2025ApJS..277...24A}. Its legend rounds this interval to
$T_0+[290,300]$~s. Using the central parameters in their table~9
without rescaling, the 100-keV reconstruction reproduces the
trace between 0.1 and 3~MeV with a maximum difference of
approximately 0.8~per~cent, within the graphical precision of
the published curve. The literal 1-keV interpretation at the
same numerical $K_1$ gives fluxes more than three orders of
magnitude smaller in this band. This supports the 100-keV
normalization for reproducing the displayed low-energy continuum,
independently of the software default. The comparison uses a
plotted model, not independent spectral measurements or the
original fit products.

This agreement does not extend to the full SBPL+Gaussian+SBPL
reference. Larger differences from the published trace remain
near the MeV feature and at higher energies; for example, the
reconstructed total is approximately 25~per~cent lower at
8.74~MeV. The low-energy continuum agreement alone therefore
does not validate the molecular ratio of 1.82 or identify the original software
configuration. That ratio remains a diagnostic of our explicitly
specified reconstruction using the central tabulated parameters.
The unfavourable outcome persists under the two normalization
interpretations tested, with the 100-keV reconstruction providing
the more permissive reference of the pair.

We therefore performed a sensitivity check in which only the
denominator was changed, using the extracted central-model trace
as the alternative reference. The same 40 archived molecular
spectra retained their normalization to the tabulated Gaussian
energy flux in 9.5--13.5~MeV and were evaluated on the same
1500-point grid over 0.1--100~MeV. All 40 spectra still exceeded
the reference at some sampled energy. The minimum maximum ratios
for Lw3o, Mw3o, Qw3o and Rh3o became approximately 3.46, 1.48,
1.60 and 1.40, respectively
(Table~\ref{tab:h2_figure8_sensitivity}). The most favourable
case remained Rh3o with $\tau_{\rm ref}=0.5$ and
$\sigma_j/E_j=0.12$, but its maximum shifted from approximately
8.74 to 19.1~MeV. Thus the mismatch persists under this change
of reference, although its magnitude is reference-dependent.
This comparison uses a graphical representation of the published
model, not independent spectral measurements. It neither
propagates observational uncertainties nor provides a statistical
exclusion, and it remains conditional on the original molecular
normalization.

\begin{table}[tp]
\centering

\small
\begin{adjustbox}{max width=\textwidth}
\begin{tabular}{lrrr}
\hline
Template & $\min R_{\max}^{\rm tabulated}$ &
$\min R_{\max}^{\rm trace}$ & $E_{\max}^{\rm trace}$ (MeV) \\
\hline
Lw3o & 3.48 & 3.46 & 1.48 \\
Mw3o & 2.00 & 1.48 & 18.8 \\
Qw3o & 2.11 & 1.60 & 8.78 \\
Rh3o & 1.82 & 1.40 & 19.1 \\
\hline
\end{tabular}
\end{adjustbox}
\caption{Minimum $R_{\max}$ for the molecular configurations
in Table~\ref{tab:h2_multiline}, using the tabulated-parameter
reconstruction with a 100-keV first-SBPL pivot or the published
model trace of Axelsson et al.~\cite{2025ApJS..277...24A}.
Molecular normalizations are unchanged.
Minima are evaluated independently over ten configurations
per template; all use $\sigma_j/E_j=0.12$.
$E_{\max}^{\rm trace}$ is the energy of the maximum ratio
for the configuration minimizing $R_{\max}^{\rm trace}$.
These ratios are diagnostics, not statistical exclusions.}
\label{tab:h2_figure8_sensitivity}
\end{table}

\begin{figure}[tp]
\centering
\includegraphics[width=0.86\textwidth]{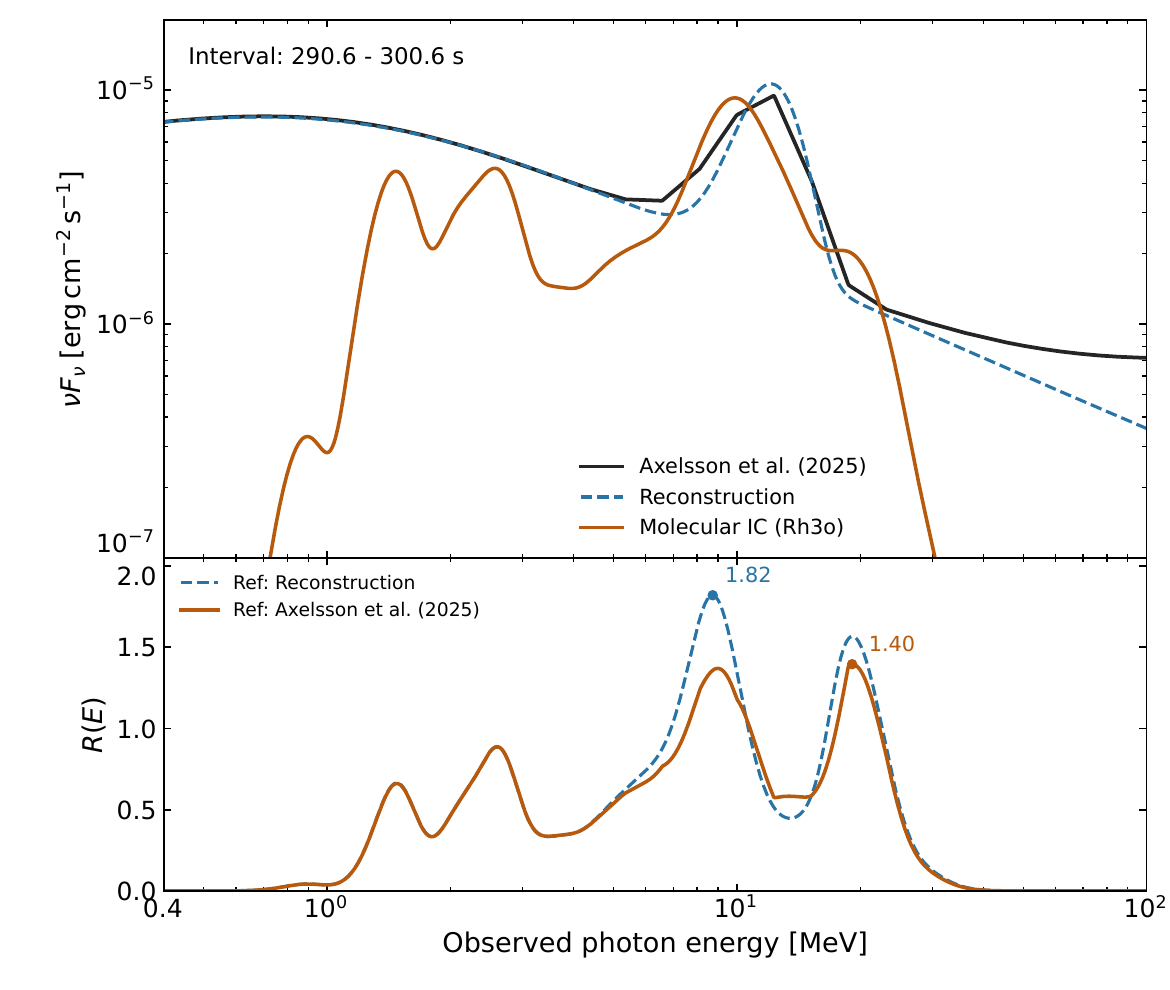}
\caption{Spectral consistency test for the most favourable molecular
configuration, Rh3o with $\tau_{\rm ref}=0.5$ and
$\sigma_j/E_j=0.12$, in the indicated interval relative to $T_0$.
Upper panel: molecular IC spectrum (orange), total central
reconstruction with a 100-keV first-SBPL pivot (dashed blue), and
published central-model trace (black). The references use table~9
and figure~8 of Axelsson et al.~\cite{2025ApJS..277...24A}, respectively.
The molecular normalization to 9.5--13.5~MeV is unchanged.
Lower panel: $R(E)=F_{E,{\rm mol}}(E)/F_{E,{\rm ref}}(E)$,
the molecular-to-reference spectral flux ratio, with the denominator
identified in the legend. The maxima are 1.82 near 8.74~MeV for the reconstruction
and 1.40 near 19.1~MeV for the published trace.
These central-model comparisons diagnose a conditional mismatch,
not a statistical exclusion.}
\label{fig:h2_multiline_consistency}
\end{figure}

Figure~\ref{fig:h2_multiline_consistency} displays the most favourable
tested configuration against both central spectral references.
The molecular spectrum and its absolute normalization are identical
in the two comparisons; the lower panel isolates the effect of
changing the reference on the apparent excess.

The Rh3o template gives the smallest maximum excess among the tested
configurations and an approximate description of the MeV feature,
but still exceeds both adopted central spectral references at some
energies. The significance of these differences has not been established.
A coupled treatment of the illumination geometry, seed-photon transfer
and electron evolution is needed to determine how the predicted spectrum
and its discrepancies change; the present estimates do not establish
that such refinements would improve the agreement.

\FloatBarrier
\section{Conclusions}
\label{sec:summary}

We investigated IC scattering of molecular and atomic
line photons as an interpretation of the evolving MeV
spectral feature in GRB~221009A. The candidates included
H$_2$ transitions, hydrogen Ly$\alpha$ and H$\alpha$,
Mg~XII, Si~XIV, S~XVI and iron K-shell lines.
For the molecular scenario, we also considered the
UV radiation responsible for fluorescence and used
contemporaneous LAT observations to assess the consistency
of the predicted IC emission from this UV field.

For lateral illumination by the isolated $2.122\,\mu{\rm m}$
H$_2$ transition, narrow lognormal electron distributions
reproduced the photon centroids and standard deviations
of the published MeV components in two observing intervals.
This calibration established agreement with the spectral
moments, rather than a fit to the complete line profiles.
The four-interval extension allowed a decreasing sequence
of electron Lorentz factors within the published marginal
parameter ranges, although the central calibrated values
were not monotonic. Since each interval was calibrated
independently, these results did not establish the
dynamical evolution of a single electron population.
The joint confidence of the decreasing sequence could
not be determined without parameter covariances.

A separate test with monoenergetic electrons included
the accompanying transitions from four molecular
templates, five external optical depths and two prescribed
widths. The Rh3o template gave the smallest maximum excess among the
tested configurations and an approximate description of the MeV feature. After normalization to the
MeV-component energy flux in 9.5--13.5~MeV, however,
all tested configurations exceeded the adopted central
spectral references at some energies. The smallest
maximum model-to-reference ratios were approximately
1.82 near 8.74~MeV for the tabulated-parameter reconstruction
and 1.40 near 19.1~MeV for the published model trace.
The excess persisted under both tested continuum-normalization
interpretations and the graphical-reference comparison,
with molecular normalizations held fixed.
These residual discrepancies were conditional on the
adopted references and scattering prescriptions.
Their statistical significance was not quantified,
as the comparison did not jointly account for the
continuum, observational uncertainties and instrumental
response.

For the adopted lateral geometry, our calculations
also predicted IC scattering of Lyman--Werner photons
into the sub-GeV band by the electron populations
calibrated to the isolated H$_2$ line. The predicted
photon centroids were approximately 210 and 161~MeV
in the earlier and later intervals, with standard
deviations of 35 and 55~MeV, respectively.
The component strengths depended on the local
seed-density ratio $q=u_{\rm UV}/u_{\rm H_2}$,
the incident spectrum and the illumination geometry.
We compared predictions for the illustrative values
$q=0.03$--0.10 with reconstructed contemporaneous LAT
spectral references to assess possible overproduction
of high-energy emission. These calculations provided
predictions for an associated IC--UV contribution,
but the comparison established neither a distinct
LAT component nor a statistical upper limit on $q$.

We found that the atomic candidates were kinematically
viable for producing MeV photons in the reference
configurations, with recoil becoming important for
the X-ray transitions. Host-galaxy recombination
emission and the marginal late-time Mg~XII feature
provided observational motivation, but did not determine
the local photon densities during the MeV episode.
We likewise did not establish the corresponding seed
fields for Si~XIV, S~XVI and iron K-shell transitions.
A physically viable interpretation would require
a source supplying sufficient photons with the appropriate
timing and angular distribution, together with a
consistent treatment of the accompanying emission.

The observed gamma-ray precursor motivated a possible preparation
mechanism for the molecular seed field, although its UV spectrum was
unmeasured. For an assumed UV budget, improved interception reduced the
ideal energy-storage deficit substantially, but the compact examples
still fell short of the required reference-line photon density by
factors of approximately $4$--$9$, even with ideal excitation and internal
escape. Under the single-use photon bound, their minimum deficits were
approximately $87$--$1.7\times10^2$. These limits did not demonstrate
a fluorescent source supplying the required field. The concentration
of excited molecules introduced additional conditions on collisional
survival and IR escape, whose evaluation depended on the material
thickness, level populations and radiation transfer.

The required accompanying molecular field also gave initial electron
cooling times comparable to the crossing times. A constant-field
estimate without acceleration yielded an approximately $35$~per~cent
decrease in Lorentz factor during transit, using only the reference
line and ten rotational transitions. Maintaining the prescribed
electron population would therefore require energy input during the
interaction or a transport calculation. Furthermore, central-precursor
illumination did not establish the lateral UV field adopted for the
sub-GeV prediction; residual UV would also require suitable redirection
or another source geometry.

The isolated-line calibration, molecular-template comparison and
analytical source estimates addressed different conditions and did not
constitute a single self-consistent configuration. They established
kinematic possibilities and conditional spectral and energetic
constraints, without demonstrating a physical origin of the MeV feature
or excluding line-seed IC scattering in general. A coupled treatment of
the seed source, radiation transfer and electron evolution would
determine how the spectral discrepancies change and whether the photon
supply, timing and illumination requirements can be satisfied together.

\section*{Data Availability}

The published spectral parameters used in this work are available
in Axelsson et al.~\cite{2025ApJS..277...24A}. The molecular-line templates of
Draine and Bertoldi~\cite{1996ApJ...468..269D} are publicly available at
\url{https://www.astro.princeton.edu/~draine/pdr.html}.
The calculation scripts, numerical spectra and spectral-reference
reconstruction files supporting the results presented here are
available from the corresponding author on reasonable request.
This work does not include a new reduction of GBM or LAT event data.

\appendix

\section{Four-interval inverse-Compton moment calibration}
\label{app:four_intervals}

We extend the isolated-line calculation of
Section~\ref{sec:h2_uv_calculation} to four observing intervals,
$T_0+[290.6,300.6]$, $[300.6,325.6]$, $[325.6,340.6]$ and
$[340.6,355.6]$~s, using the Gaussian-component parameters in
table~9 of Axelsson et al.~\cite{2025ApJS..277...24A}.
The lognormal electron distribution, lateral illumination,
observer alignment and differential Klein--Nishina kernel are
unchanged. The host-frame seed fields are the $0.58428$-eV
H$_2$ line and a prescribed UV field with uniform energy density
per unit energy between 11.2 and 13.6~eV, with $z=0.151$.
Each interval is calibrated independently; this extension is not
a time-dependent particle-transport calculation.

For the broad later components, the Gaussian location parameter
must be distinguished from the mean photon energy on the physical
domain. We adopt
$G(E)=F_{\rm G}\exp[-(E-\mu)^2/(2\sigma^2)]/(\sqrt{2\pi}\sigma)$,
where $E$ is the observed energy and $F_{\rm G}$ is the integral
over the full mathematical real axis. The published differential
amplitude is retained for $E>0$, without redistributing the
negative-energy tail. With $a=\mu/\sigma$,
$P=\Phi(a)$ and $\lambda=\phi(a)/P$, where $\phi$ and $\Phi$
are the standard normal density and cumulative distribution,
the calibration targets are
\begin{eqnarray}
 \overline E_+ &=& \mu+\sigma\lambda, \nonumber\\
 \sigma_+^2 &=& \sigma^2(1-a\lambda-\lambda^2), \nonumber\\
 {\cal F}_+ &=& c_{\rm MeV}F_{\rm G}
              [\mu P+\sigma\phi(a)].
 \label{eq:four_positive_moments}    
\end{eqnarray}

The physical photon flux is $F_{\rm G}P$.
We retain the tabulated central normalizations
$F_{\rm G}=(0.198,0.254,0.290,0.230)\,{\rm ph\,cm^{-2}\,s^{-1}}$,
in temporal order. The common IC normalization is fixed by
\({\cal F}_+\) for the H$_2$ component and then retained for UV
scattering, with $q=u_{\rm UV}/u_{\rm H_2}$ defined as in
Section~\ref{sec:h2_uv_calculation}.

\begingroup
\makeatletter
\def\fps@table{htbp}
\makeatother
\begin{table}[tp]
\centering

\small
\setlength{\tabcolsep}{4.3pt}
\begin{adjustbox}{max width=\textwidth}
\begin{tabular}{lrrrrrrrr}
\hline
Interval & $\mu$ & $\sigma$ & $\overline E_+$ & $\sigma_+$
 & $\gamma_0$ & $s_\gamma$ & $\overline E_{\rm UV}$ & $\sigma_{\rm UV}$ \\
(s) & (MeV) & (MeV) & (MeV) & (MeV) & ($10^3$) & & (MeV) & (MeV) \\
\hline
290.6--300.6 & 11.5 & 2.0 & 11.5 & 2.00 & 3.30 & 0.0867 & 210 & 35.4 \\
300.6--325.6 & 8.8 & 3.2 & 8.83 & 3.16 & 2.70 & 0.174 & 161 & 54.4 \\
325.6--340.6 & 3.0 & 4.0 & 4.56 & 2.98 & 1.62 & 0.300 & 84.3 & 51.6 \\
340.6--355.6 & 5.8 & 3.5 & 6.17 & 3.16 & 2.08 & 0.242 & 114 & 54.6 \\
\hline
\end{tabular}
\end{adjustbox}
\caption{Four-interval IC calibration using positive-energy photon
moments. Times are relative to $T_0$ and all photon energies are observed.}
\label{tab:four_intervals}

\smallskip
\begin{minipage}{0.98\textwidth}
\footnotesize
$\mu$ and $\sigma$ are the published central Gaussian parameters;
$\overline E_+$ and $\sigma_+$ are the adopted physical moments.
The UV columns are predictions from the same calibrated electrons.
Reported derived quantities are rounded; the archived calculations retain
full numerical precision. Widths are photon standard deviations, not
parameter uncertainties.
\end{minipage}
\end{table}
\endgroup

Table~\ref{tab:four_intervals} gives the calibrated moments and
electron parameters. The distinction is largest in the third
interval: $\mu=3$~MeV and $\sigma=4$~MeV give
$\overline E_+=4.56$~MeV, because approximately $22.7$~per cent
of the mathematical Gaussian integral lies at $E<0$.
The corresponding fraction is $4.9$~per cent in the fourth interval.
The same convention is applied to all four intervals for consistency.
It shifts the second-interval median from the original
$\gamma_0\simeq2.69\times10^3$ to $2.70\times10^3$; the first interval is essentially
unchanged. A separate control reproduces the original two-interval
parameters and spectra when their original moment and normalization
prescriptions are restored.

\begin{figure}[tp]
\centering
\includegraphics[width=0.98\textwidth]{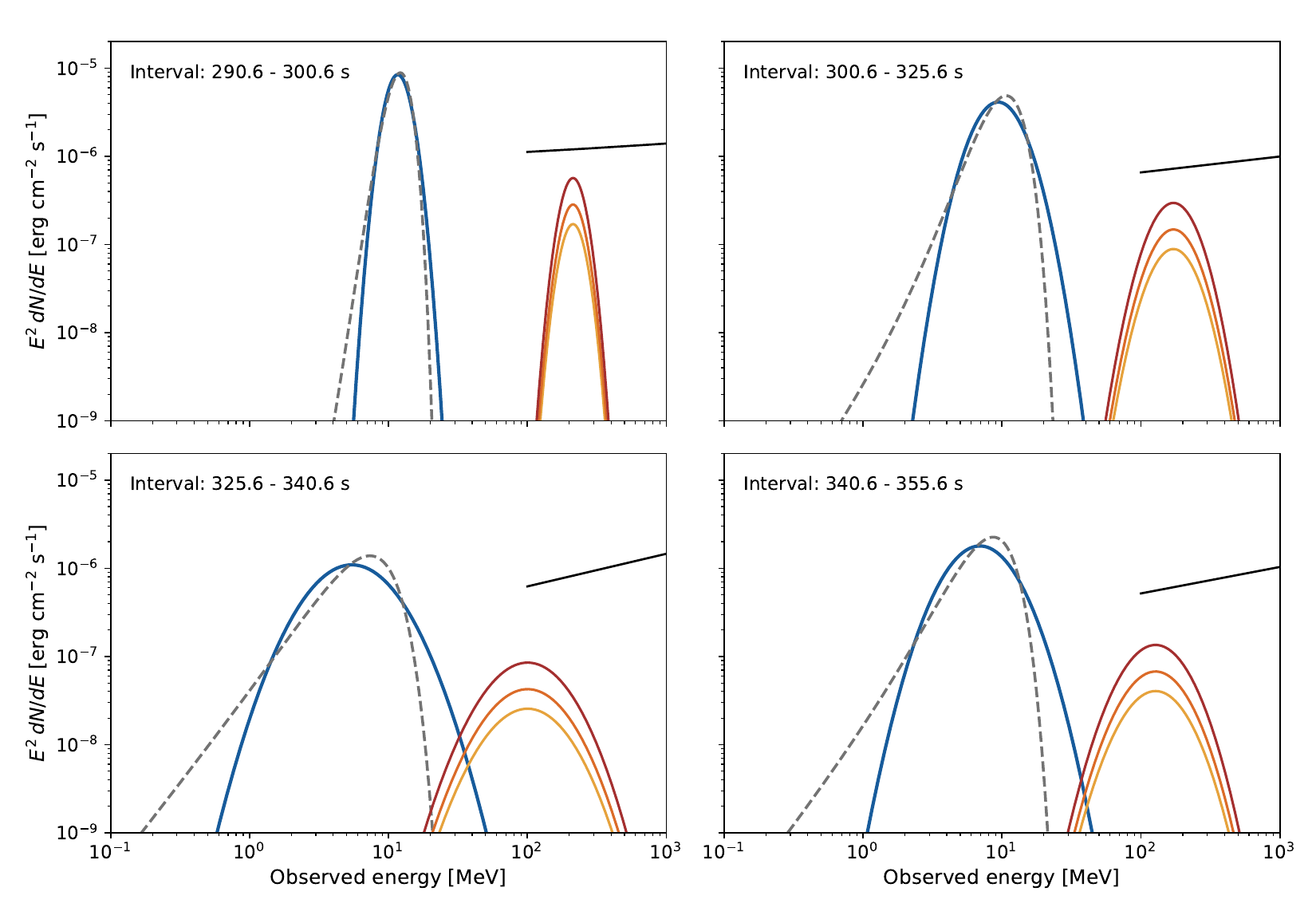}
\caption{Four-interval extension of Figure~\ref{fig:ic_components}, using the positive-energy photon moments and energy flux defined in equation~\eqref{eq:four_positive_moments}. Blue: IC–H$_2$; grey dashed: published Gaussian reference. Gold, orange and red: IC–UV for the illustrative local ratios $q=0.03$, $0.05$ and $0.10$, respectively. Black: reconstructed published LAT power-law references. The GBM continuum and accompanying molecular transitions are omitted; these components do not constitute a joint fit to detector counts.}
\label{fig:ic_four_intervals}
\end{figure}

Figure~\ref{fig:ic_four_intervals} compares the resulting IC
spectra with the published Gaussian references.
The LAT curves are reconstructed central power-law models from
table~3 of Axelsson et al.~\cite{2025ApJS..277...24A}, normalized over
$0.1$--$100$~GeV and displayed over $100$--$1000$~MeV.
They are neither independent spectral points nor identified IC--UV
components. The values $q=0.03$, $0.05$ and $0.10$ remain illustrative;
this comparison supplies no detection significance or statistical
upper limit on $q$. No event-level LAT likelihood or new exposure
calculation is introduced.

The central calibrated sequence is not monotonic, with
$\gamma_0\simeq(3.30,2.70,1.62,2.08)\times10^3$.
A bounded sensitivity check combines the lower, central and upper
published values of $\mu$ and $\sigma$, giving nine configurations
per interval while keeping $F_{\rm G}$ fixed.
For example, using $\mu_3=6$~MeV and $\sigma_3=4$~MeV within the
published marginal ranges, and retaining the other intervals at
their central values, gives a decreasing sequence
$\gamma_0\simeq(3.30,2.70,2.11,2.08)\times10^3$.
These combinations are not samples from a joint posterior;
without the parameter covariances, they cannot be assigned a
joint confidence level or interpreted as evidence for cooling.

Matching the moments does not ensure agreement with the complete
line profile, as the spectral wings in
Figure~\ref{fig:ic_four_intervals} illustrate.
Neither the central sequence nor the sensitivity examples constitute
a fit to the reported inverse-time centroid evolution.
A dynamical interpretation would require the electron distribution,
seed geometry and photon-flux weighting to evolve consistently
within the observing intervals.
This isolated-line extension does not resolve the accompanying
molecular-line mismatch or establish the physical supply of seeds.

\clearpage
\bibliographystyle{JHEP}
\bibliography{references_ADS}

\end{document}